\documentclass{mn2e}
\usepackage{epsfig,natbib2,natbibmnfix}
\usepackage{amsmath}
\usepackage{mathrsfs}
\usepackage{subfigure}
\usepackage{url}

\newcommand{\be}{\begin{equation}}
\newcommand{\ee}{\end{equation}}

\def\ltsima{$\; \buildrel < \over \sim \;$}
\def\simlt{\lower.5ex\hbox{\ltsima}}
\def\gtsima{$\; \buildrel > \over \sim \;$}
\def\simgt{\lower.5ex\hbox{\gtsima}}
\newcommand\sgra{Sgr~A$^*$}

\newcommand\Mdot{\dot{M}}
\def\del#1{{}}

\def\msun{{\,{\rm M}_\odot}}
\newcommand\mbh{{\,{\rm M}_{\rm bh}}}
\def\rsun{{\,R_\odot}}

\title[Simulations of the formation of stellar discs in the GC]{Simulations of the formation of stellar discs in the Galactic centre via cloud-cloud collisions}

\author[Alexander Hobbs and Sergei Nayakshin] {\parbox{18cm}{Alexander Hobbs$^{*}$
    and Sergei Nayakshin}\vspace{0.3cm}\\
\noindent Dept. of Physics \& Astronomy, University of Leicester, Leicester, LE1 7RH, UK}

\begin{document}

\maketitle

\begin{abstract}
Young massive stars in the central parsec of our Galaxy are best explained by
star formation within at least one, and possibly two, massive self-gravitating
gaseous discs. With help of numerical simulations, we here consider whether
the observed population of young stars could have originated from a large
angle collision of two massive gaseous clouds at $R \simeq 1$ pc from \sgra.
In all the simulations performed, the post-collision gas flow forms an inner,
nearly circular gaseous disc and one or two eccentric outer filaments,
consistent with the observations. Furthermore, the radial stellar mass
distribution is always very steep, $\Sigma_* \propto R^{-2}$, again consistent
with the observations. All of our simulations produce discs that are warped by between
30$^\circ$ to 60$^\circ$, in accordance with the most recent observations. The 3D
velocity structure of the stellar distribution is sensitive to initial
conditions (e.g., the impact parameter of the clouds) and gas cooling
details. For example, the runs in which the inner disc is fed intermittently
with material possessing fluctuating angular momentum result in multiple
stellar discs with different orbital orientations, contradicting the observed
data.  In all the cases the amount of gas accreted by our inner boundary
condition is large, enough to allow \sgra\ to radiate near its Eddington limit
over $\sim 10^5$ years. This suggests that a refined model would have
physically larger clouds (or a cloud and a disc such as the circumnuclear disc) colliding at
a distance of a few parsecs rather than 1 parsec as in our simulations.
\end{abstract}

\renewcommand{\thefootnote}{\fnsymbol{footnote}}
\footnotetext[1]{E-mail: {\tt alexander.hobbs@astro.le.ac.uk }}

\section{Introduction}

The power output of the central parsec of our Galaxy is dominated by young (aged 
$6 \pm 1$ million years) massive ``He-I'' stars \citep{KrabbeEtal95}. Many of these
stars are located in a well-defined thin disc structure that rotates clockwise as seen
on the sky \citep{Levin03,GenzelEtal03,PaumardEtal06}. The remaining stars can be
classified as a second more diffuse disc rotating counter clockwise
\citep{GenzelEtal03,PaumardEtal06} although the statistical significance of
this feature is disputed \citep{LuEtal06}. Stars appear to be on more
eccentric orbits in the second disc \citep{PaumardEtal06,LuEtal06}. The
bright He-I stars seem to be excluded from the central arcsecond ($1''\approx
0.04$ pc when viewed from the $\approx 8$ kpc distance to the Galactic Centre
(GC)), which instead contains at least a dozen less massive but still quite
young B-type stars \citep{GhezEtal03,GhezEtal05}. These stars (commonly
referred to as S-stars in the Galactic Centre community) appear to be on
rather eccentric orbits distributed without any notable preference to a planar
configuration \citep[e.g.,][]{EisenhauerEtal05,LuEtal06}. At present it is not
clear if the S-stars are physically related to the more massive
``disc''-stars, as the initial orbits of the S-stars could have been very
different from their present day ones due to resonant relaxation processes
\citep{HopmanAlexander06}. Thus, in this paper we shall concentrate on the
puzzle of the young massive stars found outside the inner arcsecond as these
present us with a cleaner laboratory environment.

\cite{Gerhard01} suggested that the stars may have been formed at a distance
of tens of parsecs in a massive star cluster. The cluster's orbit would then
decay through dynamical friction with the background stars, and eventually the
stars would form a disc populated by stars on eccentric orbits
\citep[e.g.,][]{Hansen03,Kim04,Gurkan05,LevinEtal05}. However, infrared
observations find no young massive stars outside the central $\sim 0.5$ pc
\citep{PaumardEtal06} which should be present in this model. In addition,
Chandra X-ray limits \citep{NS05} constrain the number of X-ray emitting Young
Stellar Objects to well below what is expected in the cluster disruption
scenario.

An alternative model is the in-situ model in which stars form inside a
self-gravitating gaseous disc \citep[e.g.,][]{Levin03,NC05,PaumardEtal06}. In
fact, theorists had expected gaseous massive discs around supermassive black
holes to form stars or planets
\citep[e.g.,][]{Paczynski78,Kolykhalov80,Shlosman89,Collin99,Gammie01,Goodman03}
long before the properties of the young stars in the GC became
known. Recently, \cite{NayakshinEtal07} have numerically simulated the
fragmentation process of a geometrically thin gaseous disc of mass $\simgt
10^4 \msun$ for conditions appropriate for our GC (albeit with a rather
simple cooling prescription) and found a top-heavy mass function for the stars
formed there. \cite{AlexanderEtAl08} extended numerical studies to the
fragmentation of eccentric accretion disks.

However, the in-situ model has not so far addressed in detail the origin of
the gaseous discs themselves, although similar discs are believed to exist in
AGN and quasars. Indeed, sub-parsec massive gaseous discs are invoked as means
of feeding supermassive black holes' immense appetite for gaseous fuel
\citep[e.g.][]{Frank02}. Such discs are actually observed as the sites of
maser emission in some of the nearby galactic centres \citep{Miyoshi95}. But
\sgra\ is currently not a member of the AGN club, let alone the even more
powerful and elitist quasar community. Indeed, \sgra\ bolometric luminosity is
around $\sim 10^{-9}$ of its Eddington limit
\citep[e.g.,][]{Narayan02,Baganoff03a}. The amount of ionised gas currently present in
the inner parsec is estimated at perhaps a mere hundred solar masses
\citep{Paumard04}. There are also tight observational constraints
\citep{Falcke97,Cuadra04} on the presence of an optically thin disc on smaller
scales (say 0.01 pc or less).

We believe that the requisite gaseous discs could not have been assembled
by viscous transport of angular momentum as is expected to be
the case in other systems \citep{Shakura73,Frank02}.  The viscous time scale of a
thin, marginally self-gravitating disc can be estimated as $t_{\rm visc} =
(\mbh/M_{\rm disc})^2 (\alpha \Omega)^{-1}$, where $\mbh \approx 3.5 \times
10^6\msun$ and $M_{\rm disc}$ are masses of the blackhole and the disc,
respectively, and $\Omega$ is the Keplerian angular frequency. For
$\alpha=0.1$ and $M_{\rm disc} \approx 10^4 \msun$ (reasons for this
choice of mass are discussed in \cite{NayakshinEtal06}), $t_{\rm visc} \sim 3 \times
10^7$ years at $r = 0.03$ parsec, the inner edge of the discs
\citep{PaumardEtal06}, and $10^9$ years at 0.3 parsec. These times are very
long compared with the age of the stellar systems. In fact, the gaseous discs
would have to evolve even faster as the two stellar systems are co-eval within
one million years \citep{PaumardEtal06}.

It is also not clear that one should expect any sort of a viscous
quasi-steady state for the gas in our Galactic Centre. There is no strong evidence for
other similar star formation events in the central parsec within the last
$\sim 10^8$ years. Therefore, the one-off star formation event appears to be
best explained by a one-off deposition of gas within the central parsec. There
are several ways in which this could have happened, e.g. a Giant Molecular Cloud
(GMC) with a sub-parsec impact parameter (in relation to \sgra) could have
self-collided and become partially bound to the central parsec
\citep[e.g.,][]{NC05}. Alternatively, a GMC could have struck the
Circumnuclear Disc (CND) located a few parsecs away from \sgra, and then
created gas streams that settled into the central parsec.

It is very hard to estimate the probability of a cloud-cloud collision in the
central part of the Galaxy. \cite{Hasegawa94} estimated the rate of GMC
collisions as one per 20 Myrs in the central 100 pc, but the estimate is
uncertain by at least an order of magnitude in either direction due to the
lack of knowledge about the size distribution of GMCs. However, there is
observational evidence that GMCs can be put on orbits significantly
different from the simple circular orbits seen in the inner Galaxy. \cite{Stolte08}
recently measured the proper motions of the Arches star cluster. This star
cluster is about 2.5 million years old and has a mass in excess of $10^4 \msun$
\citep[e.g.,][]{Figer04}. It is presently about 30 pc away from \sgra\
in projection. Its orbit must be strongly non-circular as its 3D velocity is
over 200 km/sec in the region where the circular velocity is only $\sim 110$
km/sec. Therefore, this suggests that massive GMCs can be on highly
non-circular orbits. It does not seem implausible that one of these clouds would
pass within a few parsec of \sgra.

In this paper we explore such a one-off collision event in a very simple setup
(\S \ref{sec:ic}). We allow two massive, uniform and spherical clouds on
significantly different orbits to collide with each other one parsec away from
\sgra. The resulting gas dynamics, in particular the way in which gas settles
into the inner parsec, is the focus of our effort here. We find that the
collision forms streams of gas with varying angular momentum, both in
magnitude and direction. Parts of these streams collide and coalesce to form a
disc, and remaining parts form one or more orbiting filaments. As the gas
cools, it becomes self-gravitating and stars are born, usually in both the
disc and the filaments. This overall picture is discussed in \S
\ref{sec:overall}. A reader mainly interested in a comparison between the
simulations and observations may find sections \ref{sec:radial},
\ref{sec:accretion} and \ref{sec:discussion} most relevant\footnote{For movies of the
simulations the reader is directed to \url{http://www.astro.le.ac.uk/~aph11/movies.html}}. 
In \S \ref{sec:obs} we argue that our simulation results suggest that the location
of the original impact was somewhat further away from \sgra\ than we assumed
here, e.g., perhaps a few parsecs.

\section{Numerical methods}\label{sec:numerics}

The numerical approach and the code used in this paper is the same as in
\cite{NayakshinEtal07} with only slight modifications. We employ
GADGET-2, a smoothed particle hydrodynamic (SPH)/N-body code
\citep{Springel05}. The Newtonian N-body gravitational interactions of
particles in the code are calculated via a tree algorithm. Artificial
viscosity is used to resolve shocks in the gas. Gas cools according to $du/dt
= - u/t_{\rm cool}(R)$, where cooling time depends on radius as
\begin{equation}
t_{\rm cool}(R) = \beta  t_{\rm dyn}(R)
\label{beta}
\end{equation}
\noindent where $t_{\rm dyn} = 1/\Omega$ and $\Omega = (GM_{\rm
bh}/R^{3})^{1/2}$ is the Keplerian angular velocity, and $\beta$ is a
dimensionless parameter. This approach is motivated by simulations of marginally
stable self-gravitating gaseous discs, where $\beta$ is expected to be of the
order of a few \citep{Gammie01,Rice05}. In the problem at hand, $\beta$ would
most likely be much smaller than this during the collision phase of the clouds as gas
can heat up to high temperatures (see \S \ref{sec:obs}), but it could be much
larger than unity when gas cools into a geometrically thin disc or a thin
filament. In the framework of our simplified approach, and given numerical
limitations, we only consider two values for $\beta$.  The $\beta=1$ tests
approximate the situation when cooling is inefficient, whereas $\beta = 0.3$
cases correspond to rapid cooling. Each of our runs used between 3 to 6 months of
integration time on 12 processors in parallel.

Gas moves in the gravitational potential of \sgra, modelled as a motionless
point mass with $M_{\rm bh} = 3.5 \times 10^{6} \msun$ \citep{Reid99} at the
origin of the coordinate system, and a much older relaxed isotropic stellar
cusp. For the latter component we use the stellar density profile derived from
near-IR adaptive optics imaging by \cite{GenzelEtal03}:

\vspace{-0.15in}

\begin{center}
\begin{gather}
\frac{\rho_{*}(R)}{\rho_{*0}} \; = \;
\begin{cases}
\left(\frac{R}{R_{\rm cusp}}\right)^{-1.4}, \hspace{0.2in} R < R_{\rm cusp}\\
\left(\frac{R}{R_{\rm cusp}}\right)^{-2}, \hspace{0.27in} R \ge R_{\rm cusp}
\end{cases}
\end{gather}
\label{rhocusp}
\end{center}

\noindent where $R_{\rm cusp}$ = 10 arcseconds ($\approx$ 0.4 pc), and $\rho_{*0}
= 1.2 \times 10^{6} \msun/pc^{3}$.

Note that the enclosed stellar cusp mass is approximately $6 \times 10^{5}
\msun$ at $R = R_{\rm cusp}$ and equals \sgra\ mass at $R \sim 1.6$ pc. We do not
include the gravitational potential of the putative stellar mass black hole
cluster around \sgra \citep[e.g.,][]{Morris93,Miralda00}, as at present its
existence and exact properties are highly uncertain
\citep{Freitag06,SchoedelEtal07,DeeganNayakshin07}.

Star formation and accretion in SPH is traditionally modelled via the ``sink
particle'' formalism \citep{Bate95}, where gravitationally contracted gas
haloes are replaced by the collisionless sink particles of same mass. We
follow the same recipe, allowing \sgra\ to accrete all gas particles that are
within $R_{\rm acc} \sim 0.1$ arcsecond (exact values vary between different
cooling parameters, cf. \S \ref{sec:ic}). This is significantly
smaller than the inner radius of the observed ``disc'' population of young
massive stars, i.e., $\sim 1''$ \citep{PaumardEtal06}.

To model star formation, new sink particles are introduced when gas density exceeds
\begin{equation}
\rho_{\rm crit} = \rho_{0} + A_{\rm col}\rho_{\rm BH}
\end{equation}
\noindent where $\rho_{0} = 1.67 \times 10^{-12}$ g cm$^{-3}$, $\rho_{\rm BH}$ is
the tidal density, and $A_{\rm col} = 50$
(this method is based on that of \cite{NayakshinEtal07}, which
found that the results are not sensitive to the exact values of $\rho_{0}$ and
$A_{\rm col}$, provided they are sufficiently large). For reasons fully
explained in \cite{NayakshinEtal07}, we distinguish between ``first cores" --
star particles with $M \leq M_{\rm core} = 0.1 \msun$, and ``stars" with $M >
M_{\rm core}$. Physically, first cores represent massive gaseous
clumps that have not yet collapsed to stellar densities, and thus have finite
sizes that are set to $R_{\rm core} = 10^{14}$~cm. First cores are allowed to
merge with each other if they pass within a distance $d < 2 R_{\rm
core}$. 

In contrast, stars are not allowed to merge, but can accrete first cores if
they pass within a distance of $R_{\rm core}$. Accretion of gas onto stars and
proto-stars is calculated using the Bondi-Hoyle formalism, with the accretion
rate given by
\begin{equation}
\Mdot = 4 \pi \rho \frac{(GM)^{2}}{(\Delta v^{2} + c_{\rm s}^{2})^{3/2}}
\end{equation}
where $M$ is the mass of the sink particle, $\rho$ is the ambient
gas density, $c_{\rm s}$ is the sound speed, and $\Delta v$ is the relative
velocity between the gas and the accreting particle. The accretion rate is
capped at the Eddington accretion rate:
\begin{equation}
{\dot M}_{\rm Edd} = \frac{4\pi m_{\rm p} R_* c}{\sigma_{\rm T}}\;.
\label{medd}
\end{equation}
where $m_{\rm p}$ is the proton mass, $c$ the speed of light and $\sigma_{\rm T}$ the
Thomson scattering cross-section. Note that this expression depends only on the size of the object, $R_*$. For
first cores, $R_* = R_{\rm core}$, which yields a very high accretion rate
limit of almost a solar mass per year for $R_{\rm core} = 10^{14}$ cm. For
stars, we use the observational results of \cite{1991Ap&SS.181..313D,
1998ARep...42..793G}:
\begin{eqnarray}
\frac{R}{\rsun} & = & 1.09 \left(\frac{M}{\msun}\right)^{0.969}\;\;{\rm for}\;\;   M < 1.52 \msun, \\
\frac{R}{\rsun} & = & 1.29 \left(\frac{M}{\msun}\right)^{0.6035} \;\;{\rm for}\;\;  M > 1.52 \msun .
\end{eqnarray}
For $R=\rsun$, this yields an Eddington accretion rate limit of $\sim 5 \times
10^{-4} \msun$~year$^{-1}$.  The actual gas particles that are then accreted
by the sink particle are chosen from its neighbours via the stochastic SPH
method of \cite{Springel05}. The mass and momentum of the SPH particles
accreted are added to that of the respective sink particle.

Finally, we shall acknowledge that this plausible star formation {\em
prescription} cannot possibly do justice to the full complexity of the physical star
formation process. Therefore, the stellar mass function obtained in our
simulations should not be trusted in detail. However, our focus is on the dynamics
of gas as it settles into a disc or a filamentary structure. When star
formation does occur, the gas temperature is very much smaller than the virial
temperature in the simulated region ($\simgt 10^6 K$). Gas orbits thus
determine the resulting stellar orbits which we compare to the observed
ones. We believe that this important aspect of our simulations is reliably
modelled.

The set of units used in the simulations is $M_{u} = 3.5 \times 10^{6} \msun$,
the mass of \sgra \citep[e.g.,][]{Schoedel02}, $R_{u} = 1.2 \times 10^{17}$ cm
$\approx 0.04$ pc, equal to 1'' when viewed from the $\approx$ 8 kpc distance
to the GC, and $t_{u} = 1/\Omega(R_{u})$, the dynamical time evaluated at
$R_{u}$, approximately 60 years. We use $R$ to signify distance in physical
units and the dimensionless $r = R/R_{u}$ throughout the paper
interexchangeably.

\begin{figure*}
\begin{minipage}[h]{.48\textwidth}
\centerline{\psfig{file=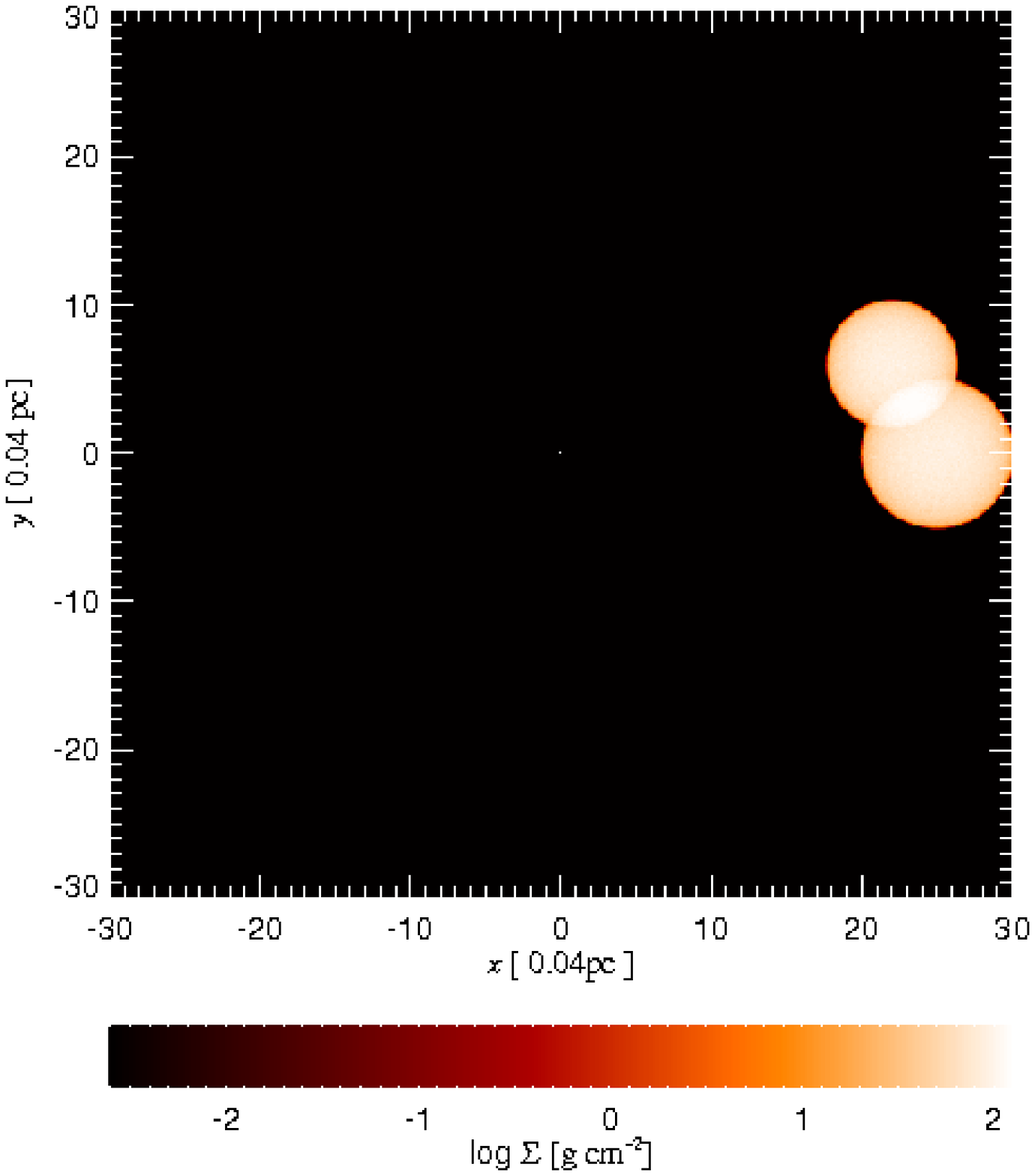,width=0.99\textwidth,angle=0}}
\end{minipage}
\begin{minipage}[h]{.48\textwidth}
\centerline{\psfig{file=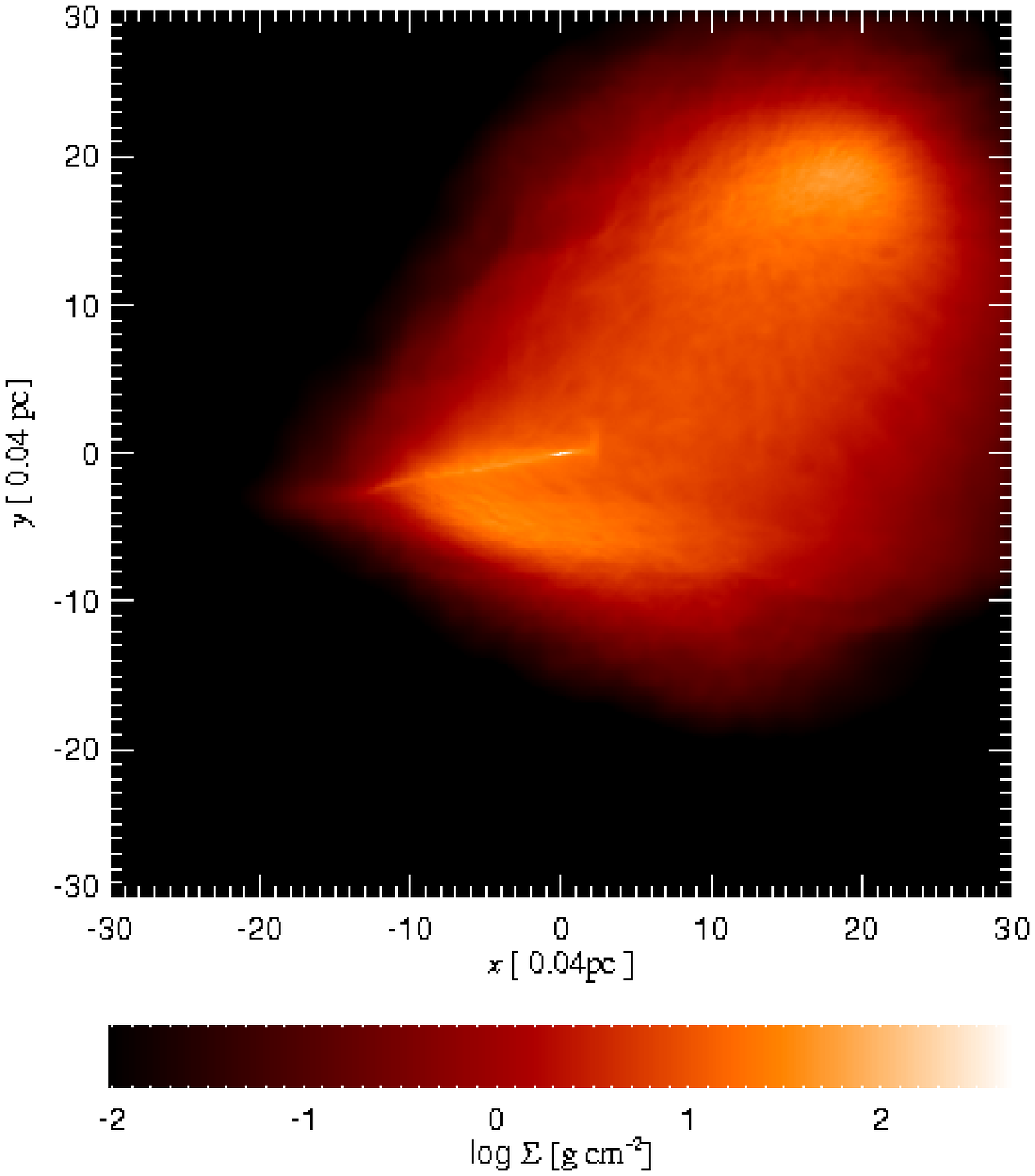,width=0.99\textwidth,angle=0}}
\end{minipage}
\begin{minipage}[h]{.48\textwidth}
\centerline{\psfig{file=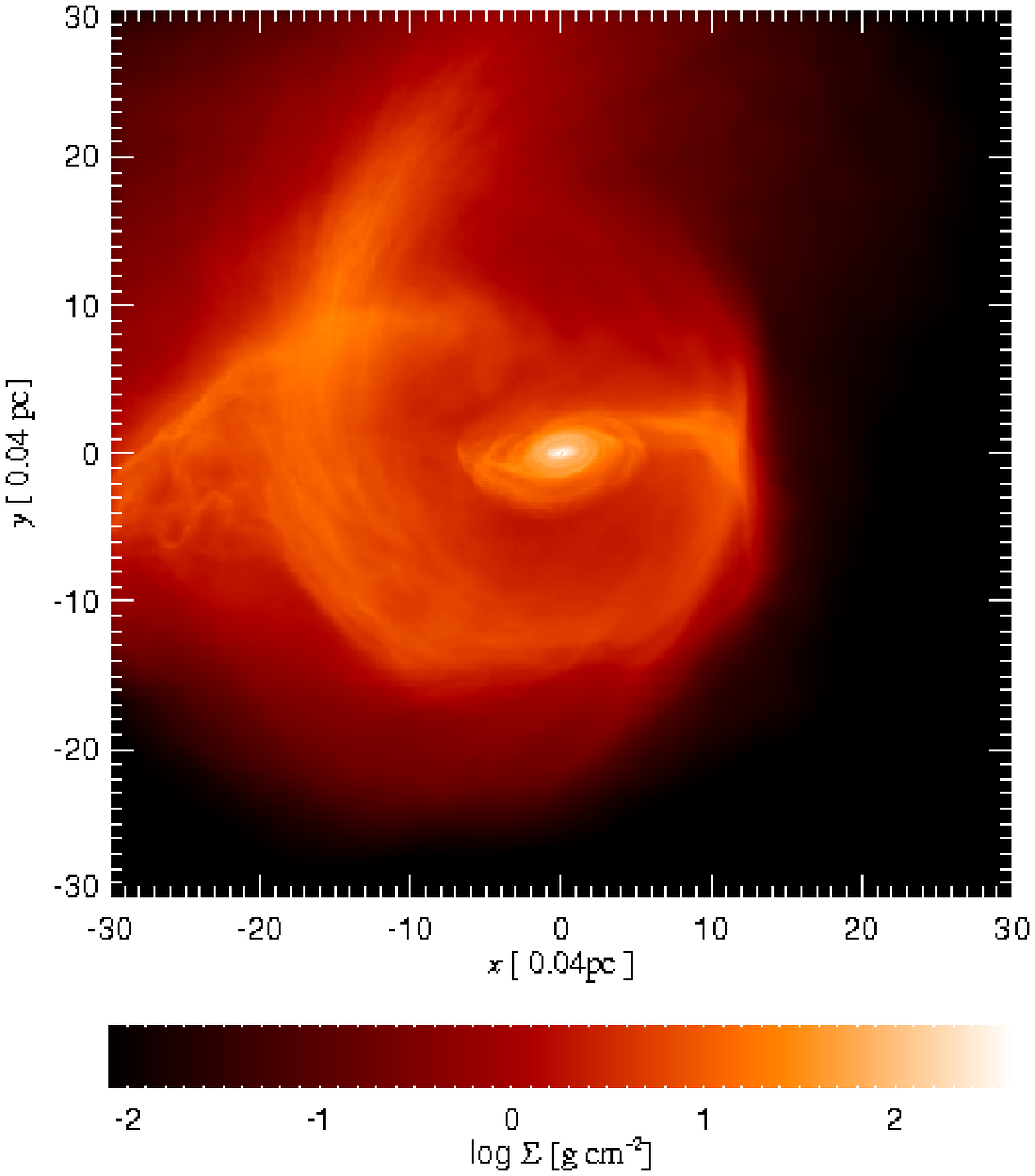,width=0.99\textwidth,angle=0}}
\end{minipage}
\begin{minipage}[h]{.48\textwidth}
\centerline{\psfig{file=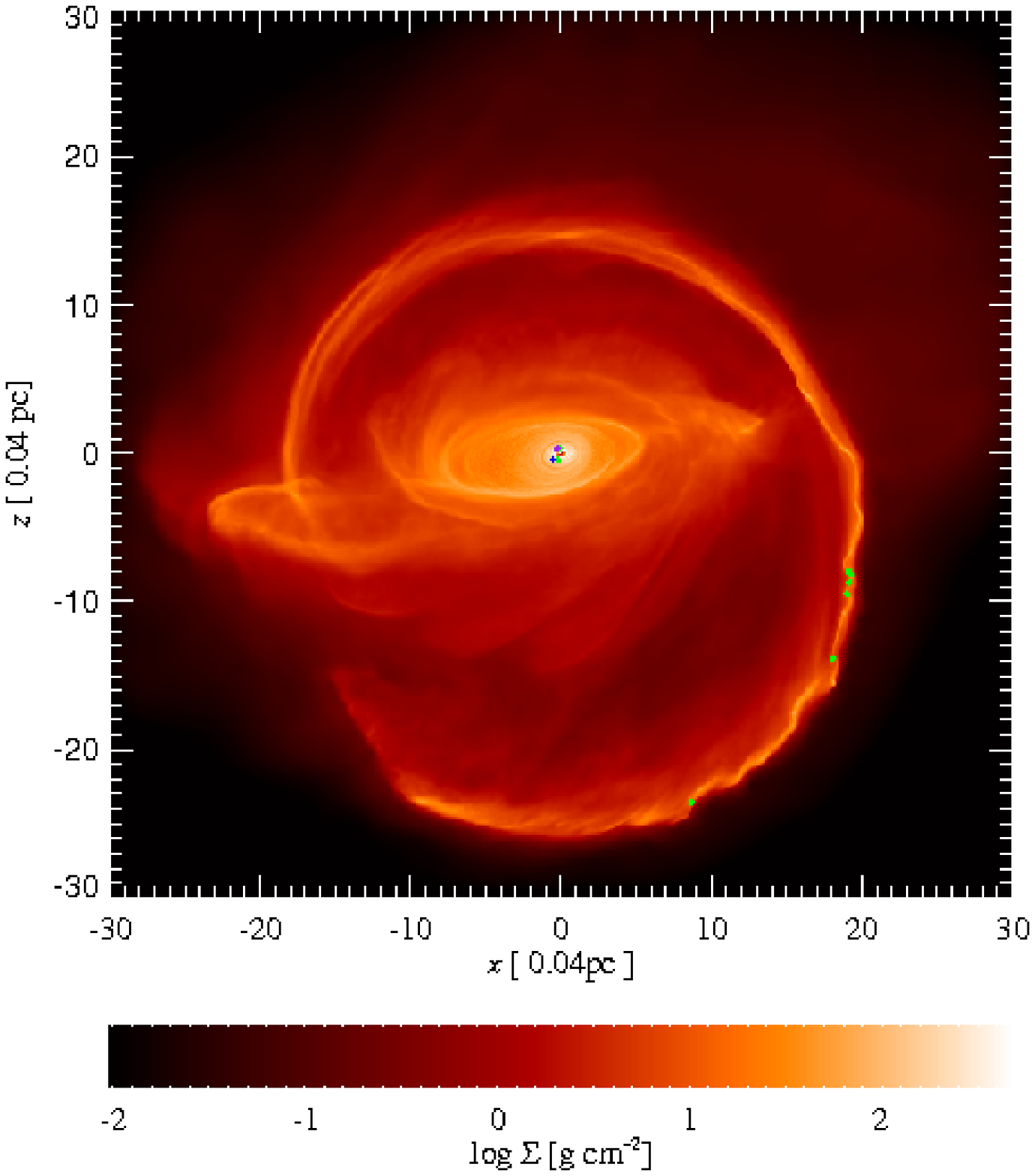,width=0.99\textwidth,angle=0}}
\end{minipage}
\caption{Gas surface density and star locations (in the bottom right panel)
  for snapshots from simulation S1 taken at times $t=0, 100, 250,$ and $1000$,
  left to right and top to bottom, respectively. \sgra\ is located at (0,0),
  and the line of sight is along the $z$-direction.}
\label{fig:S1overview}
\end{figure*}

\section{Initial conditions}\label{sec:ic}

Simulations were performed for several sets of initial conditions, all of which
comprise a collision between two gas clouds at the edge of the inner parsec of
the GC. 2,625,000 SPH particles were used in each simulation. The specific parameters for each run, labeled S1 to S6, can be found
in Table 1. Each cloud is spherically symmetric and of uniform density,
containing less than 1\% of the mass of \sgra. The clouds are composed of
molecular hydrogen, with the mean molecular weight set to $\mu = 2.46$. The
initial temperature of the clouds is set to 20K. A more complex
model could have included a turbulent velocity and density field, but for
practical purposes we limit ourselves to a smaller set of input parameters for
this first study.

We define the primary to be the larger cloud, with a radius of $R_1 = 0.2$ pc
and $M_{1} = 3.4 \times 10^{4} \msun$, whilst the secondary has a radius of
$R_2 = 0.172$ pc with $M_{2} = 2.6 \times 10^{4} \msun$.  The initial
positions of the cloud centres are the same for all the simulations, and are
given in Table 1 in dimensionless units.  The
initial density of the clouds is slightly above the tidal shear density at
their initial locations, so the clouds are marginally stable to tidal shear by
\sgra.

In all the tests the initial velocity of the primary cloud is kept fixed, and is of the
order of the Keplerian circular velocity for that radius. Orbits in a cusped
potential are however not Keplerian. The
initial specific energy and angular momentum $l$ of a particle can be
used to find the pericenter, $r_{\rm pe}$,  and the apocenter, $r_{\rm ap}$,
of the orbit. The orbit's eccentricity $e$ is then defined via
\begin{equation}
\frac{1+e}{1-e} = \frac{r_{\rm ap}}{r_{\rm pe}}\;.
\label{ecceq}
\end{equation}
In our simulations particles occasionally obtain hyperbolic orbits. For
analysis (plotting) purposes only, the eccentricities are capped at 1. In the
stellar potential used here (equation \ref{rhocusp}), the orbit of the primary cloud is
slightly eccentric, with pericentre and apocentre of 25 and 31.5 respectively,
and with eccentricity $e = 0.12$.

The initial trajectory of the secondary cloud is varied between the tests
to cover a small range of possibilities. The parameters for this trajectory in
terms of pericentre, apocentre and eccentricity are given in Table
\ref{table1}. The collision itself is highly supersonic, as is expected for
molecular gas falling into the potential well of a supermassive blackhole.

\begin{table*}
\caption{Initial conditions of the simulations presented in the paper. The
meaning of the symbols in the Table are: $\beta$ is the cooling parameter,
$\bf{r_1}$, $\bf{r_2}$, $\bf{v_1}$, $\bf{v_2}$ are the initial positions and
velocity vectors of the two clouds, respectively; $r_{\rm pe}$ and $r_{\rm
ap}$ are the pericentres and the apocentres of the two clouds; $e$ is their
orbit's eccentricity; $\theta$ is the angle between the orbital planes of the
clouds, and $b$ is the impact parameter.}
\begin{center}
\begin{tabular}{|c|c|c|c|c|c|c|c|c|c|c|c|c|c|c|}\hline
ID & $\beta$ & $\bf{r_{1}}$ & $\bf{r_{2}}$ & $\bf{v_{1}}$ & $\bf{v_{2}}$ & $|\bf{v_{1}}-\bf{v_{2}}|$ &
                                  $r_{1,\rm pe}$ & $r_{1,\rm ap}$&
                                  $r_{2,\rm pe}$ & $r_{2,\rm ap}$ & $e_{1}$ & $e_{2}$ & $\theta$ ($^{\circ}$) & $b$\\ \hline
                                  \hline
S1 & 1  & (25,0,0) & (22,6,7) & (0,0.2,0) & (0,-0.11,-0.21) & 0.37 & 25  & 31.5 & 12.8 & 29.5 & 0.12 & 0.39 & 116 & 3.8\\ \hline
S2 & 1  & (25,0,0) & (22,6,7) & (0,0.2,0) & (0,-0.21,-0.11) & 0.42 & 25  & 31.5 & 13.2 & 29.1 & 0.12 & 0.38 & 151 & 6.8\\ \hline
S3 & 0.3 & (25,0,0) & (22,6,7) & (0,0.2,0) & (0,-0.11,-0.21) & 0.37 & 25 & 31.5 & 12.8 & 29.5 & 0.12 & 0.39 & 116 & 3.8\\ \hline
S4 & 0.3 & (25,0,0) & (22,6,7) & (0,0.2,0) & (0,-0.21,-0.11) & 0.42 & 25 & 31.5 & 13.2 & 29.1 & 0.12 & 0.38 & 151 & 6.8\\ \hline
S5 & 1 & (25,0,0) & (22,6,7) & (0,0.2,0) & (0.16,-0.11,-0.21) & 0.41 &25 & 31.5 & 21.8 & 43.1 & 0.12 & 0.33 & 120 & 2.4\\ \hline
S6 & 1 & (25,0,0) & (22,6,7) & (0,0.2,0) & (0.16,-0.21,-0.11) & 0.45 &25 & 31.5 & 21.5 & 43.4 & 0.12 & 0.34 & 147 & 5.3\\ \hline
\end{tabular}
\end{center}
\label{table1}
\end{table*}

We ran tests with cooling parameter $\beta = 1$ and $\beta = 0.3$ (see
equation \ref{beta}). These values are low enough so that fragmentation would
occur if and when regions of the gas became self-gravitating
\citep{Gammie01,Rice05}. Since the faster cooling runs were expected to
require on average shorter timesteps the accretion radius for $\beta = 0.3$
was set to $r_{\rm acc} = 0.33$ whilst for $\beta = 1$ a smaller value of
$r_{\rm acc} = 0.06$ was used.

We have performed convergence tests (at the referee's request) on the numerical viscosity
parameter $\alpha$ and on $N_{\rm sph}$. We find that there is good convergence on the distribution
of gas when varying either of these parameters, although there are minor differences within the innermost part of the 
computational domain, namely the inner 0.5''-1'' around the central black hole. These differences are small and can
be explained by a different amount of gas being accreted onto the black hole when either viscosity or resolution are
changed.

\section{Simulations}

\subsection{Overall picture}\label{sec:overall}

In all of our simulations, the clouds undergo an off-centre collision at time
$t \sim 10$ in code units ($\approx 600$ years). As the cooling time is longer than the
collision time, $t_{\rm coll} \sim (R_1 + R_2)/(|\bf{v_1} - \bf{v_2}|)$, the
clouds heat up significantly and hence initially expand considerably. This thermal
expansion also modifies the velocity of the different parts of the clouds by
giving the gas velocity ``kicks''. The net result is a distribution of gas
velocities that is much broader than what one would get if the two clouds
simply stuck together, i.e. in an inelastic collision.

The collision and resultant mixing of the clouds leads to angular momentum cancellation in some
parts of the gas. Regions of gas that acquire smaller angular momenta infall
to the respective circularisation radius on the local dynamical time. A small
scale disc around the black hole is thus formed on this timescale. Regions of
the clouds that did not directly participate in the collision are affected
less and retain more of their initial angular momentum. These regions are
initially not self-gravitating, and so are sheared by the tidal field of \sgra. They later cool and result in filaments of length comparable to the initial
sizes of the clouds' orbits.

As time progresses, parts of the filaments collide with each other or with the
inner disc if the pericentres of their orbits are small enough. The inner disc therefore gains mass in an asymmetric non-planar manner, leading to the disc becoming warped and changing its orientation with
time. 

Eventually, due to gravitational collapse, high density gas clumps are formed. Sink particles are
introduced inside these clumps and are allowed to grow in mass via gas accretion
and mergers with first cores as explained in \S \ref{sec:numerics}. Star
formation takes place in all of our simulations. Although the precise
locations and the stellar mass distributions differ we always end up with stellar populations forming in two distinct locations -- in the disc around the black hole and in the outer
filament(s). Disc populations generally have top-heavy mass functions whilst filament populations always consist entirely of low-mass stars; however, as noted in \S \ref{sec:numerics}, the mass
spectrum of stars formed depends on a very simple prescription \citep[e.g., see][]{NayakshinEtal07}.  We expect the mass
spectrum of our models to change when more realistic cooling and feedback physics is
included.

\subsection{SIMULATION S1}\label{sec:s1}

\subsubsection{Gas dynamics}

Figure \ref{fig:S1overview} presents four snapshots of simulation S1 showing
gas column density and positions of the stars viewed along the $z$-axis.

The initial mass deposition into the central few arcseconds forms a
small-scale disc that changes its orientation by about $30^\circ$ with time,
stabilising by $t \sim 300$. As a result of the small impact parameter of the
cloud collision, the secondary cloud is largely destroyed and does not
collapse into a coherent filament. Instead, the diffuse gas left over from the
impact gradually accretes onto the disc that forms around the black hole,
growing it in mass and radius. The inner disc therefore ends up extending out
to $r \sim 10$ in code units. The gas left over from the primary cloud collapses into a
filament which orbits at a larger radius, separate from the disc.

\begin{figure}
\begin{minipage}[b]{.48\textwidth}
\centerline{\psfig{file=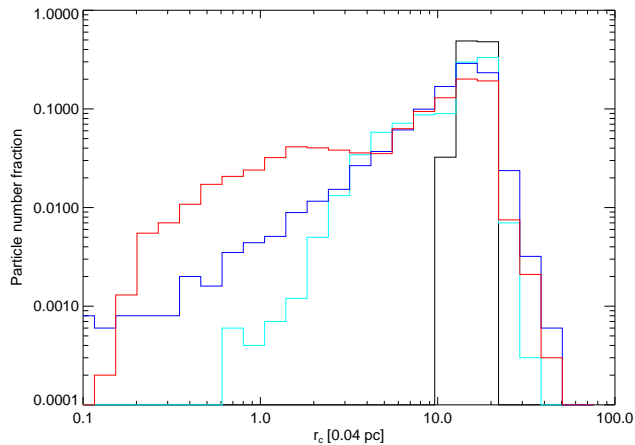,width=0.99\textwidth,angle=0}}
\end{minipage}
\caption{The distribution of SPH particles over the
  circularisation radius of their orbits. The black, cyan, blue and red
  curves correspond to times $t= 0, 25, 50, 500$, respectively. As the
  result of angular momentum cancellation due to shocks, particles ``leak'' to
  small radii, establishing there a disc that eventually forms stars.}
\label{fig:S1_rcirc_hist}
\end{figure}

It is useful to analyse the re-distribution of the gas as a result of the
collision in terms of orbital parameters, namely, angular momentum and
eccentricity. Gas with specific angular momentum $l = |\bf{r}\times \bf{v}|$ circularises at
the circularisation radius, $r_{\rm c}$, given by the relation
\begin{equation}
l^2 = G M(r_{\rm c}) r_{\rm c}\;,
\label{rc}
\end{equation}
where $M(r)$ is the total enclosed mass within radius $r$. Figure
\ref{fig:S1_rcirc_hist} shows the distribution of gas particles as a function of the
circularisation radius of their orbits as a histogram for several snapshots.
At time $t=0$, the distribution is highly peaked around the clouds' initial
circularisation radii. Due to the collision, however, a tail to small $r_{\rm c}$ appears
in the distribution very quickly. This is not due to viscous angular
momentum transport, as in thin accretion discs, but is rather due to
cancellation of oppositely directed angular momentum components in
shocks. Some of the gas particles aquire particularly small angular momenta and
hence fall to small radial separations from \sgra\ on nearly radial
trajectories. Particles with $r_{\rm c} < r_{\rm acc}$ are in general accreted by \sgra\ in our
formalism (unless their orbit is changed by interactions before they make it
inside $r_{\rm acc}$). Particles with $r_{\rm c} > r_{\rm acc}$ settle into a thin
rotationally supported disc, or into sheared out filaments (the remains of the
clouds). Note that due to the complicated geometry, some gas
particles pass through shocks several times before reaching their final
destination (i.e. the disc).

\begin{figure}
\begin{minipage}[b]{.48\textwidth}
\centerline{\psfig{file=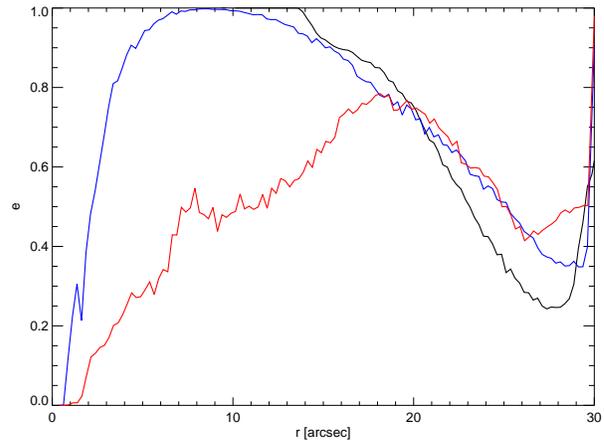,width=0.99\textwidth,angle=0}}
\end{minipage}
\caption{ Gas eccentricity defined on radial shells as a function of the shell's
  radius at different times.  Black, blue and red curves correspond to time
  $t=50$, 100 and 250. Note that initially only gas on near-plunging high
  eccentricity orbits arrives in the innermost region, but with time gas
  circularises to $e\ll 1$.}
\label{fig:ecc_radial}
\end{figure}

Figure \ref{fig:ecc_radial} shows the profiles of orbital eccentricity (defined
on radial shells) plotted for several snapshots.  These profiles again show
that soon after the collision the inner part of the computational domain is
dominated by gas on plunging -- high eccentricity -- orbits. At later times
finite net angular momentum of gas and shocks force the gas to circularise
outside the inner boundary of the simulation domain. It is notable that gas
circularises faster at smaller radii, which is naturally expected as the
orbital time is shortest there. Therefore, towards later times,
nearly circular gaseous orbits are established in the innermost few
arcseconds, whereas eccentric ones dominate at larger radii.

\subsubsection{Structures in angular momentum space}\label{sec:globus}

\begin{figure*}
\begin{minipage}[b]{.48\textwidth}
\centerline{\psfig{file=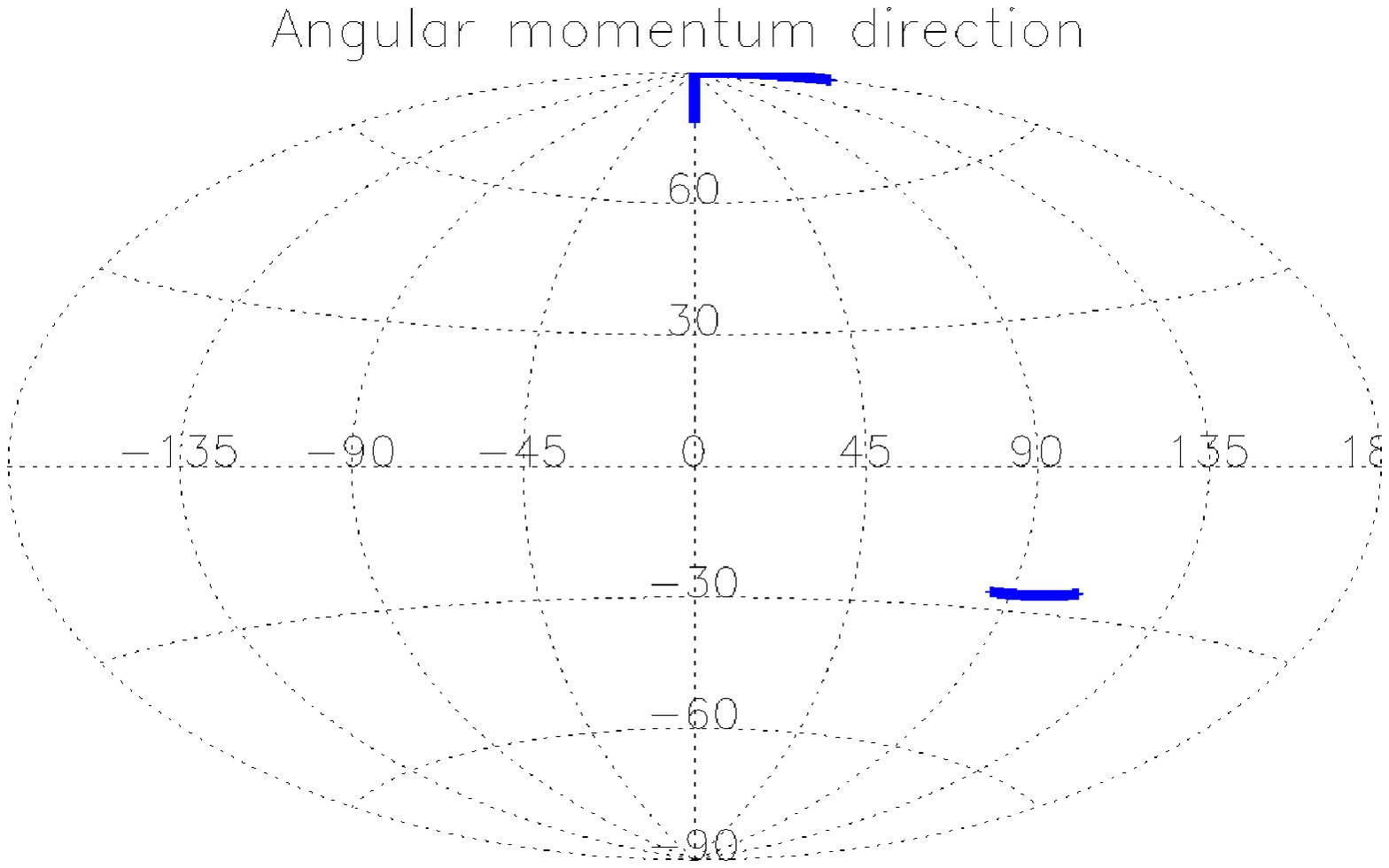,width=0.99\textwidth,angle=0}}
\end{minipage}
\begin{minipage}[b]{.48\textwidth}
\centerline{\psfig{file=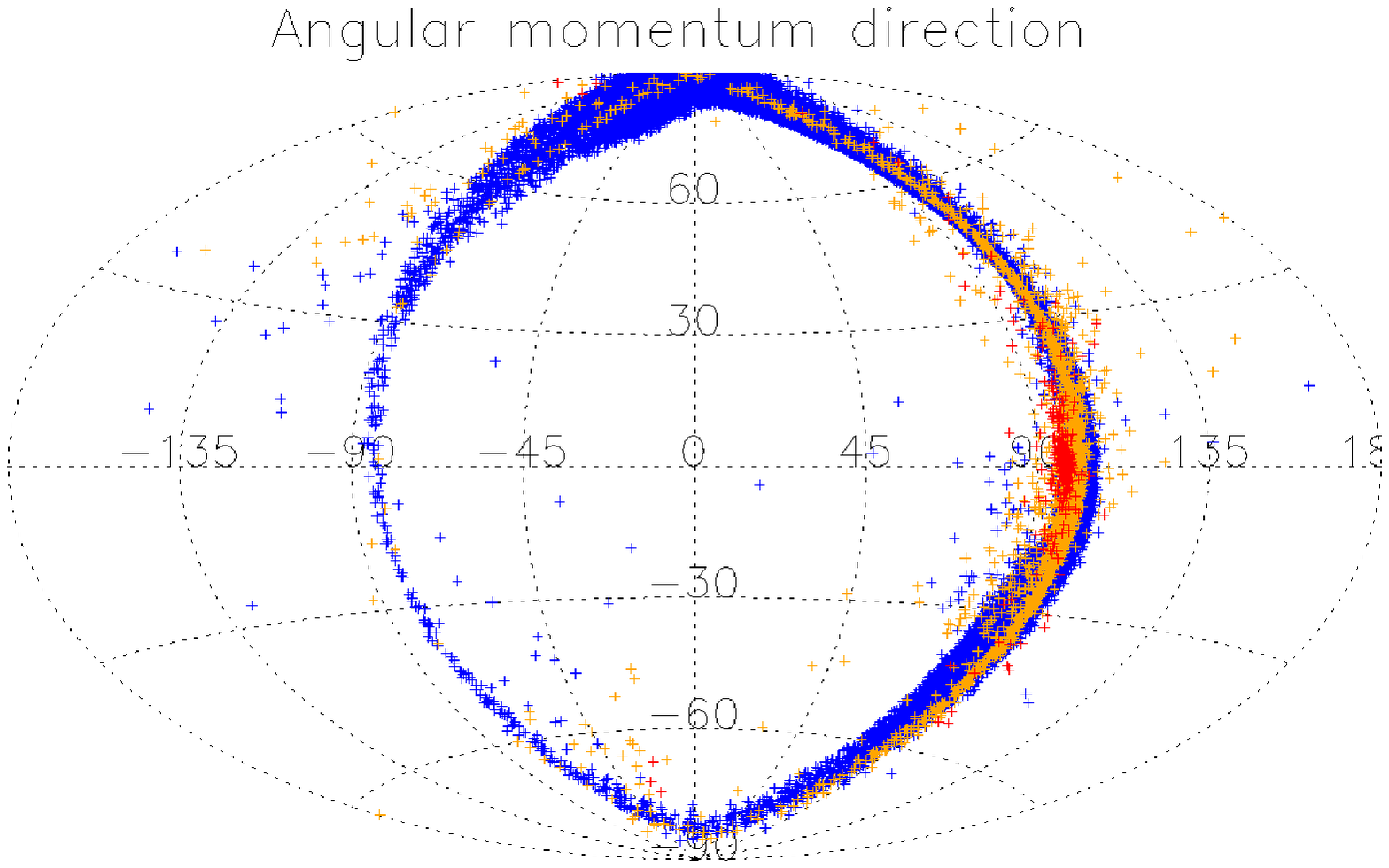,width=0.99\textwidth,angle=0}}
\end{minipage}
\begin{minipage}[b]{.48\textwidth}
\centerline{\psfig{file=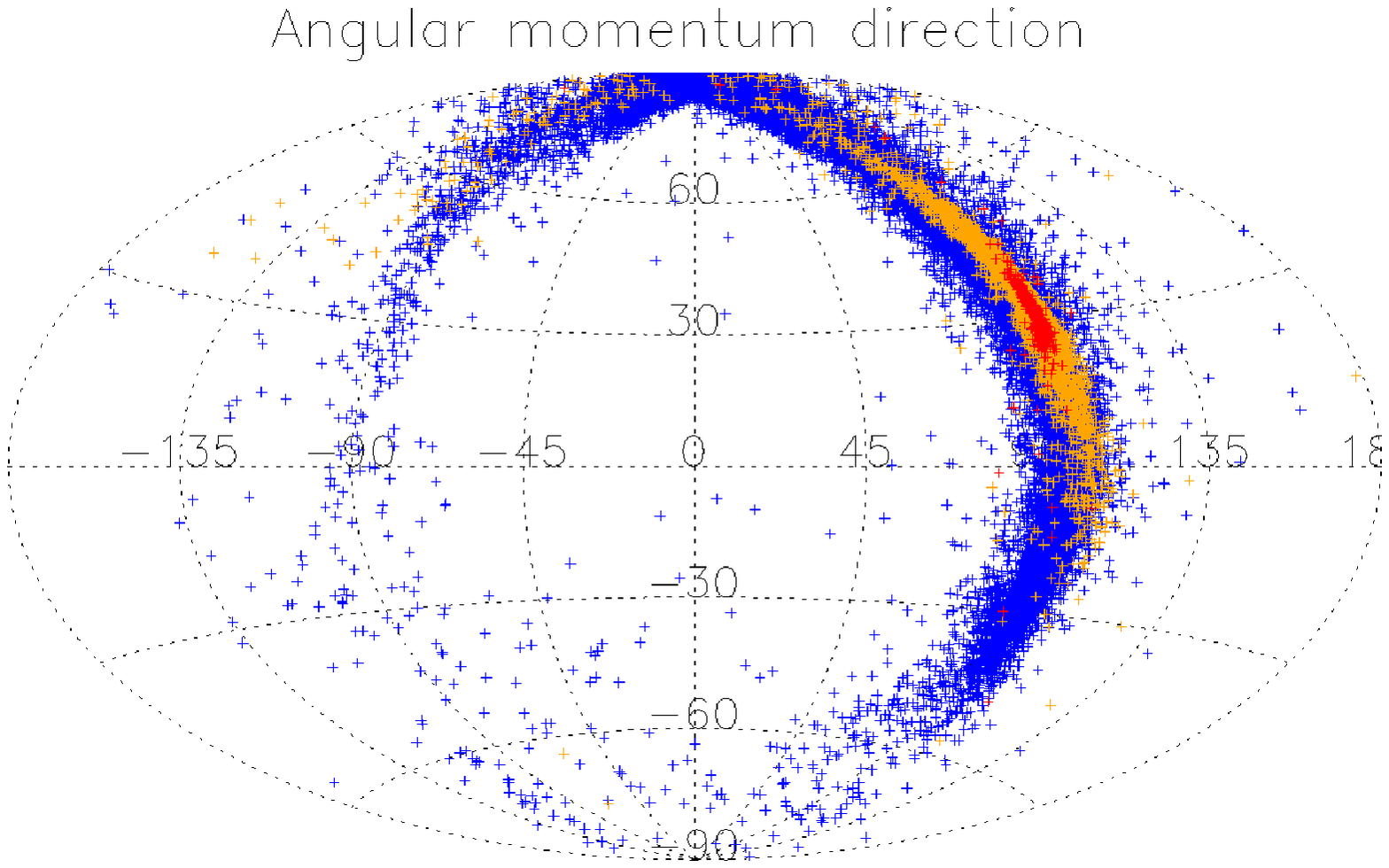,width=0.99\textwidth,angle=0}}
\end{minipage}
\begin{minipage}[b]{.48\textwidth}
\centerline{\psfig{file=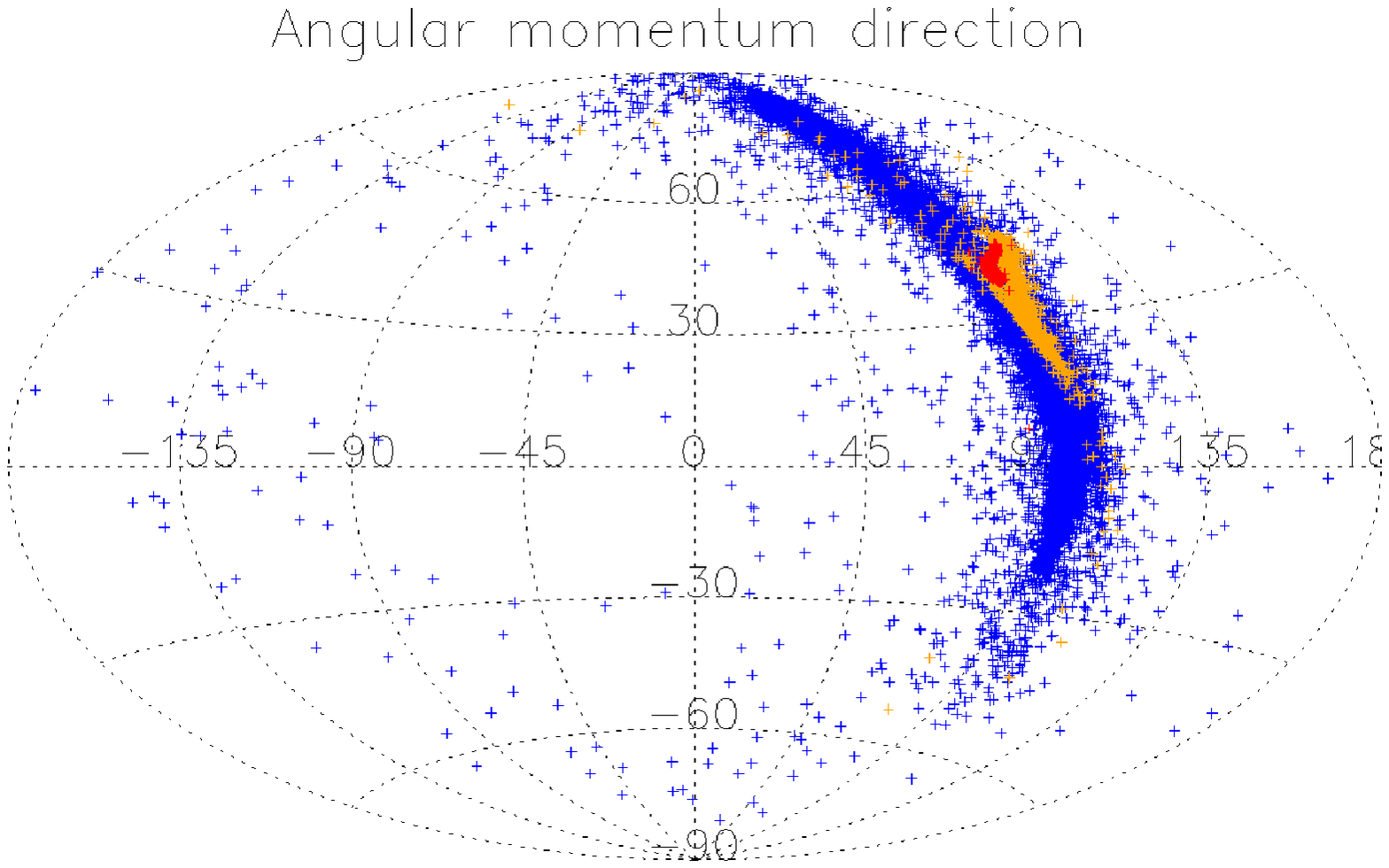,width=0.99\textwidth,angle=0}}
\end{minipage}
\caption{Angular momentum direction in terms of the azimuthal, $\theta$, and
  the polar, $\phi$, angles. The frames are shown at $t = 0,100,250,1000$. Gas
  particles are colour-coded; $R \leq 1''$ (red), $R \leq 5''$ (orange), $R >
  5''$ (blue).}
\label{fig:S1globus}
\end{figure*}

The various structures that are formed in the simulation can also be identified
by the orientation of the angular momentum vector $\bf{L}$ of a particle's orbit.
We define two angles, $\theta$ and $\phi$, so that $L_z = \vert \bf{L} \vert
\cos\theta$, where $L_z$ is the $z$-component of $\bf{L}$, and $\tan\phi =
L_y/L_x$. We then plot the values of these angles for individual SPH particles
and stars in the Aitoff projection. As the number of SPH particles is too
large to show each particle, we show only a small fraction of these, randomly
chosen from the total number. This procedure keeps the resulting maps
reasonably clear while also preserving the shape of the distribution.

The result is shown in Figure \ref{fig:S1globus} for a selection of times. At
$t=0$ all the gas particles are concentrated in the two clouds, with the
finite spread in the distribution arising as a result of all the particles
within a cloud having the same velocity whilst possessing a finite spread in
positions about the centre. The difference in the orientations of the two
clouds is almost $120^{\circ}$ (cf. Table 1).  By $t=100$, the clouds have
collided and the gas particles have experienced `kicks' in different
directions, filling the entire range in $\theta$. The $\phi$ values are
restricted to a rather narrow band running about $\phi = +90^{\circ}$ and
$\phi = -90^{\circ}$ lines. Note that this band structure forms a plane $L_x
\approx 0$ in \textbf{L}-space. This is expected since gas particles
experience the collision close to the $x$-axis, i.e., where $y/x \approx z/x
\approx 0$. Therefore, the $x$-component of the resulting angular momentum,
$L_x = {y v_z - z v_y}$ is very small.

The plots are colour-coded to show the behaviour of the gas in the inner
arcsecond (red), the inner 5 arcseconds (orange), and outside the inner 5
arcseconds (blue). Whilst the distribution just after the collision is fairly
spread out for all the three regions, we note a clear difference at later
times. In particular, the innermost disc occupies a region positioned in between the two clouds' original positions. It is also the least spread out
structure, defining a thin disc which is only slightly warped by $t=1000$. The
innermost disc orientation does evolve with time, however, as we noted already
in \S \ref{sec:overall}, since matter infall on the disc continues throughout
the simulation.

The gas coloured in orange demonstrates a greater extent of warping, and the
region outside this ($> 5''$) cannot even be classified into a single structure (cf. Figure
\ref{fig:S1overview}). Nevertheless, it is clear that the inner and outer
gas distributions are similarly oriented with respect to each other, with the
majority of the outer gas angular momentum distributed in $\theta$ and $\phi$
between the initial values of the clouds.

\subsubsection{Star formation}\label{sec:s1sf}

Stars in S1 form in both the disc and the primary filament. In the disc, stars form first in the inner arcsecond at $t \sim 900$ and later at a radius of $\sim 5-8''$ at $t \sim 1700$. In the filament, star formation is approximately co-eval with the inner arcsecond population and occurs at a radius of $\sim 15-25''$. Mass functions in the disc become top-heavy relatively quickly (as the dynamical time is short), with high-mass stars ($M \sim 100 \msun$) appearing by $t \approx 1100$, whilst stars in the filament remain entirely low-mass ($\sim 0.1 - 1 \msun$) for the duration of the simulation. The disc and filament stellar orbits are inclined to each other by $\sim 60^{\circ}$, similar to the gas orbital inclination.

All three stellar populations can be seen in Figure \ref{fig:S1S2density}; the ``inner disc'' population (within the inner $1-2''$), the ``mid-range disc'' population ($5-8''$) and the ``filament'' population. The orbits in the inner disc population are almost circular, $e \sim 0.05$, whilst stars in the mid-range disc have eccentricities of $e \sim 0.2$. Both of these are therefore in good agreement with the clockwise feature of the observations \cite{LuEtal06}, the orbits of which are only mildly eccentric. The orbit of the filament population is also eccentric, with $e \simgt 0.2$. This is in reasonable agreement with the corresponding counter-clockwise feature of the observations, although the eccentricity of the latter is considered to be somewhat higher with $e \sim 0.8$ \citep{PaumardEtal06}.

\subsection{SIMULATION S2}\label{sec:s2}

\subsubsection{Gas dynamics}\label{sec:s2dynamics}

The simulation S2 is set up identically to S1 except for the initial velocity of
the secondary cloud (cf. Table 1). As the result, the angle between the
orbital plane of the two clouds is greater, $\theta \sim 150^{\circ}$, and the
collision between the clouds is more grazing (i.e. the impact parameter $b$ is
greater).

\begin{figure*}
\begin{minipage}[b]{.48\textwidth}
\centerline{\psfig{file=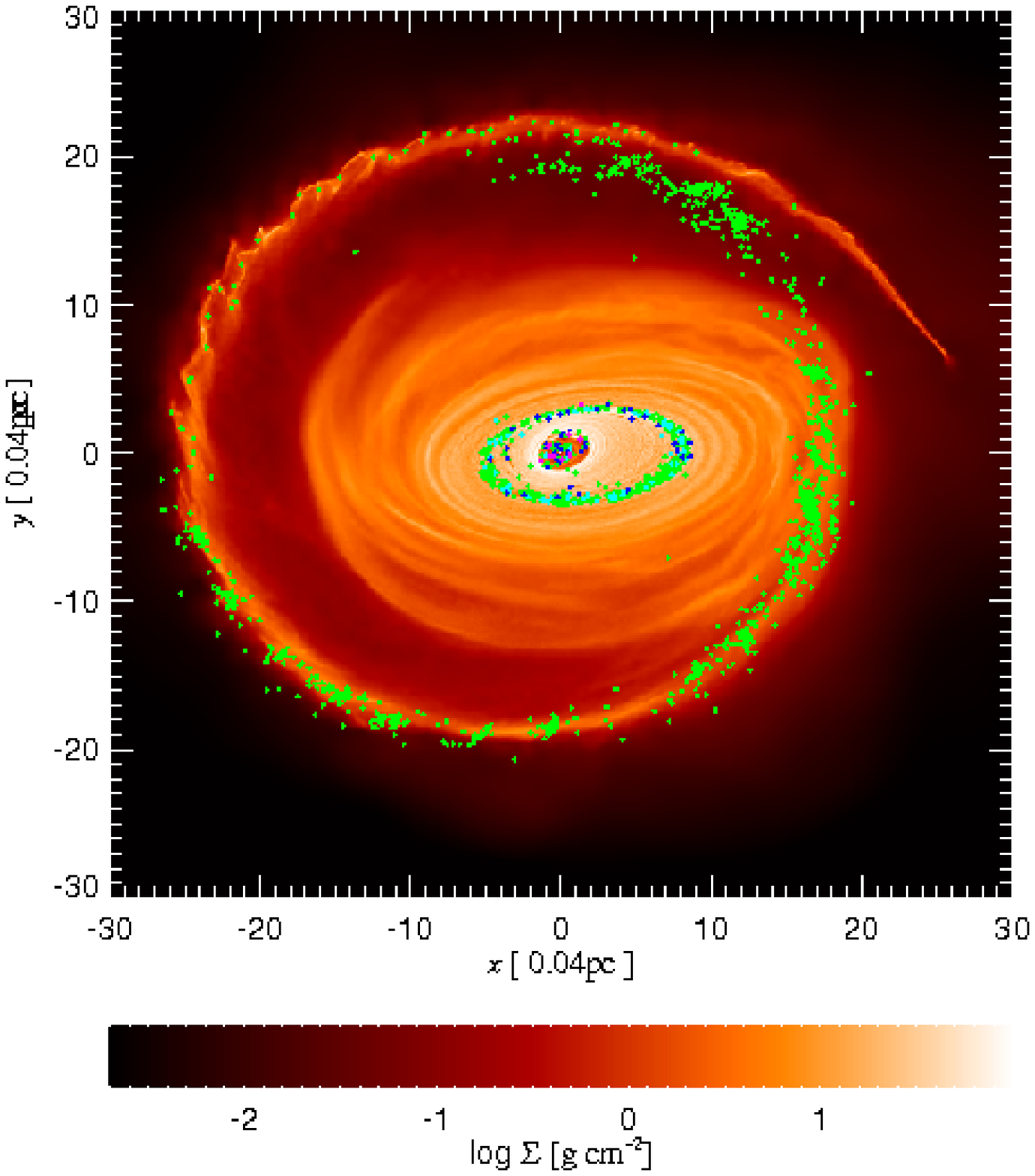,width=0.99\textwidth,angle=0}}
\end{minipage}
\begin{minipage}[b]{.48\textwidth}
\centerline{\psfig{file=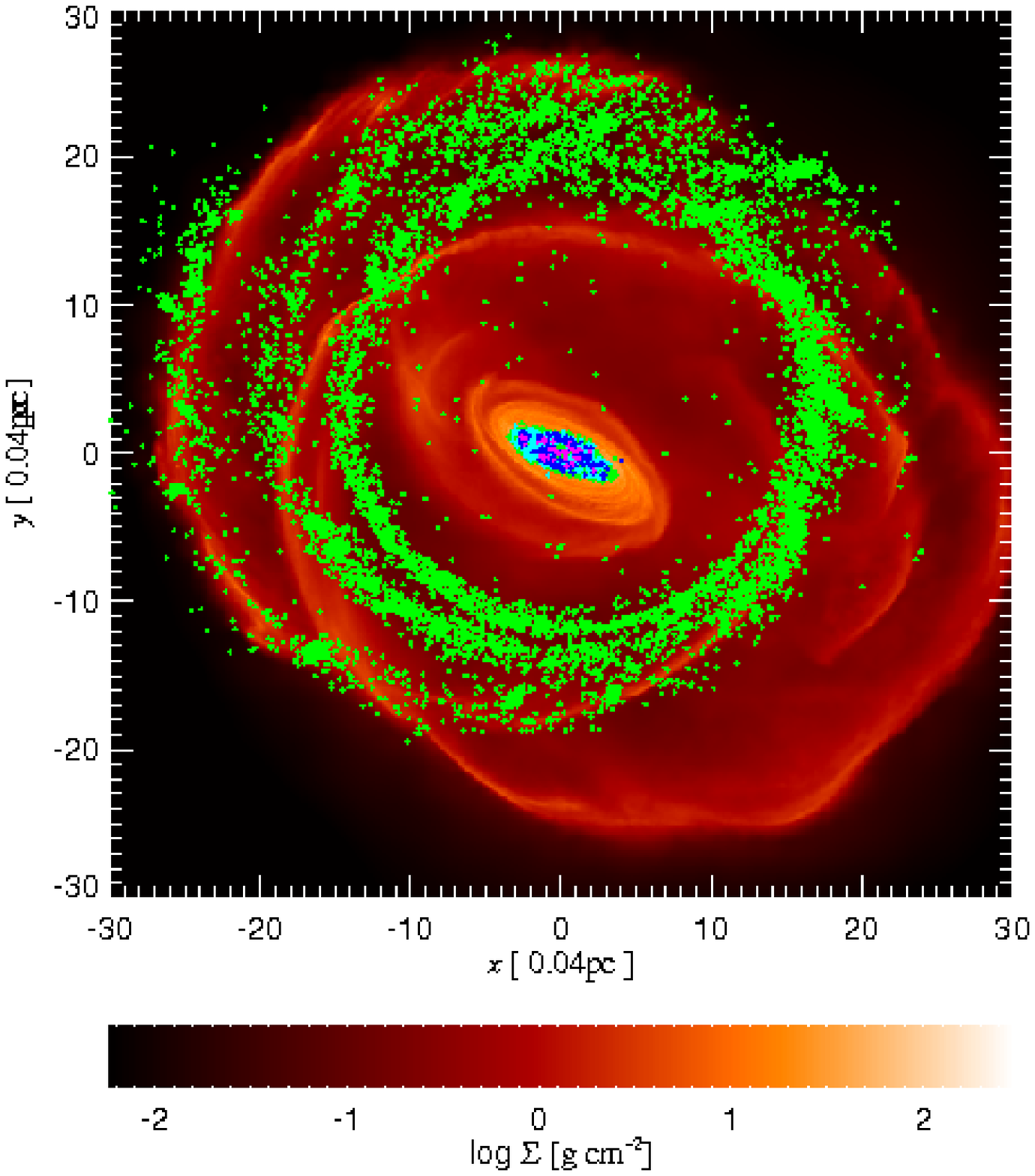,width=0.99\textwidth,angle=0}}
\end{minipage}
\caption{Projected gas densities and stellar positions for simulations S1
  (left) and S2 (right) at $t=1955$. Stars are shown by coloured symbols as
  following: green $M_* = 0.1 - 1 \msun$, cyan $M_* = 1 - 10 \msun$, blue $M_*
  > 10 \msun$, and magenta $M_* > 150 \msun$. Note the survival of the
  secondary filament in S2. The line of sight is along the $z$-direction.}
\label{fig:S1S2density}
\end{figure*}

Smaller fractions of the clouds'
mass are therefore involved in a bodily collision, and as a result the two
clouds survive the initial collision better. Indeed, comparing the S1 and
S2 simulations in Figure \ref{fig:S1S2density}, one sees that in the latter
simulation most of the secondary cloud actually does survive the collision. This
cloud forms a clockwise filament which was largely absent in simulation
S1. Both S2 filaments contain a greater amount of mass than their S1 counterpart.

\begin{figure*}
\begin{minipage}[b]{.48\textwidth}
\centerline{\psfig{file=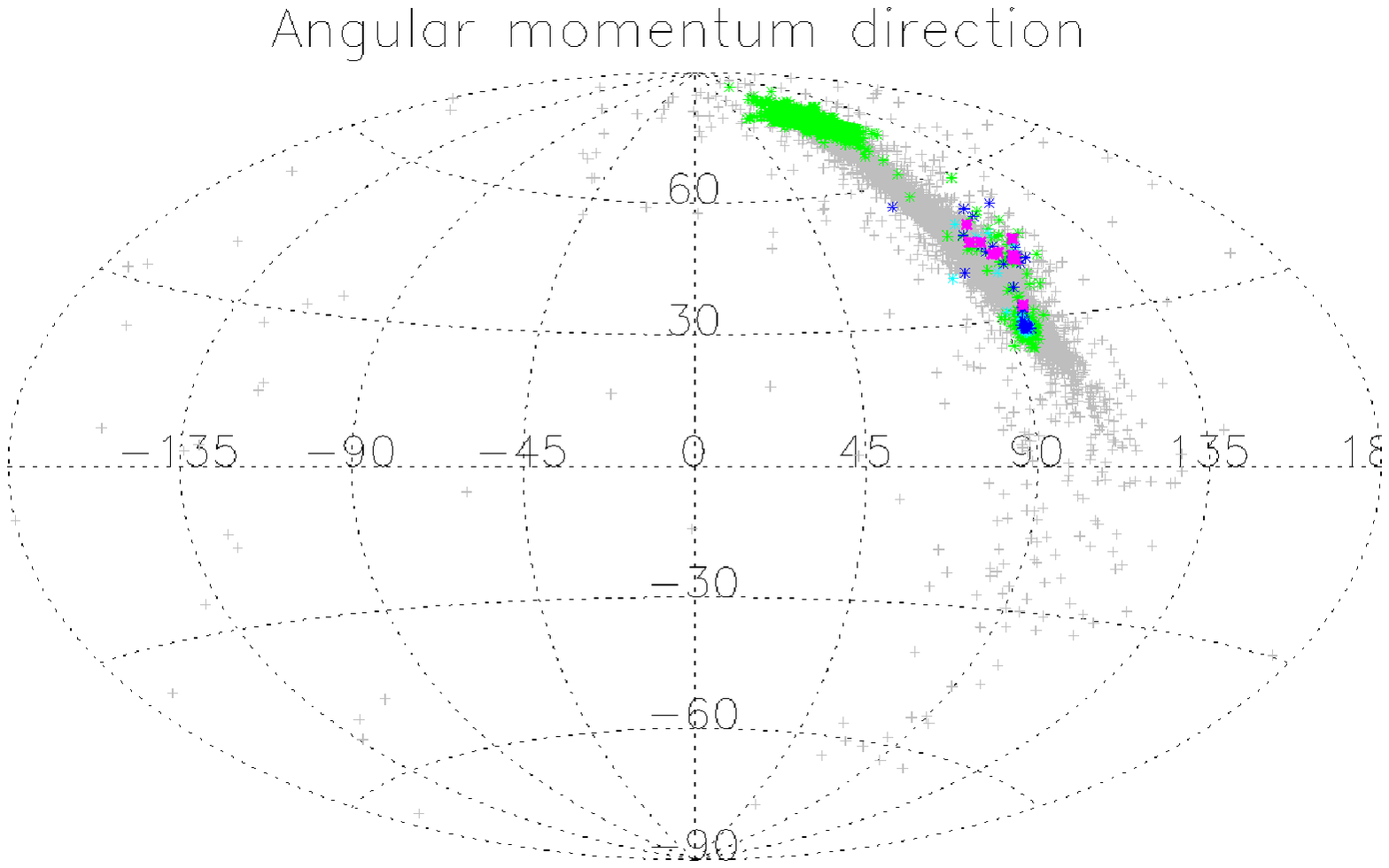,width=0.99\textwidth,angle=0}}
\end{minipage}
\begin{minipage}[b]{.48\textwidth}
\centerline{\psfig{file=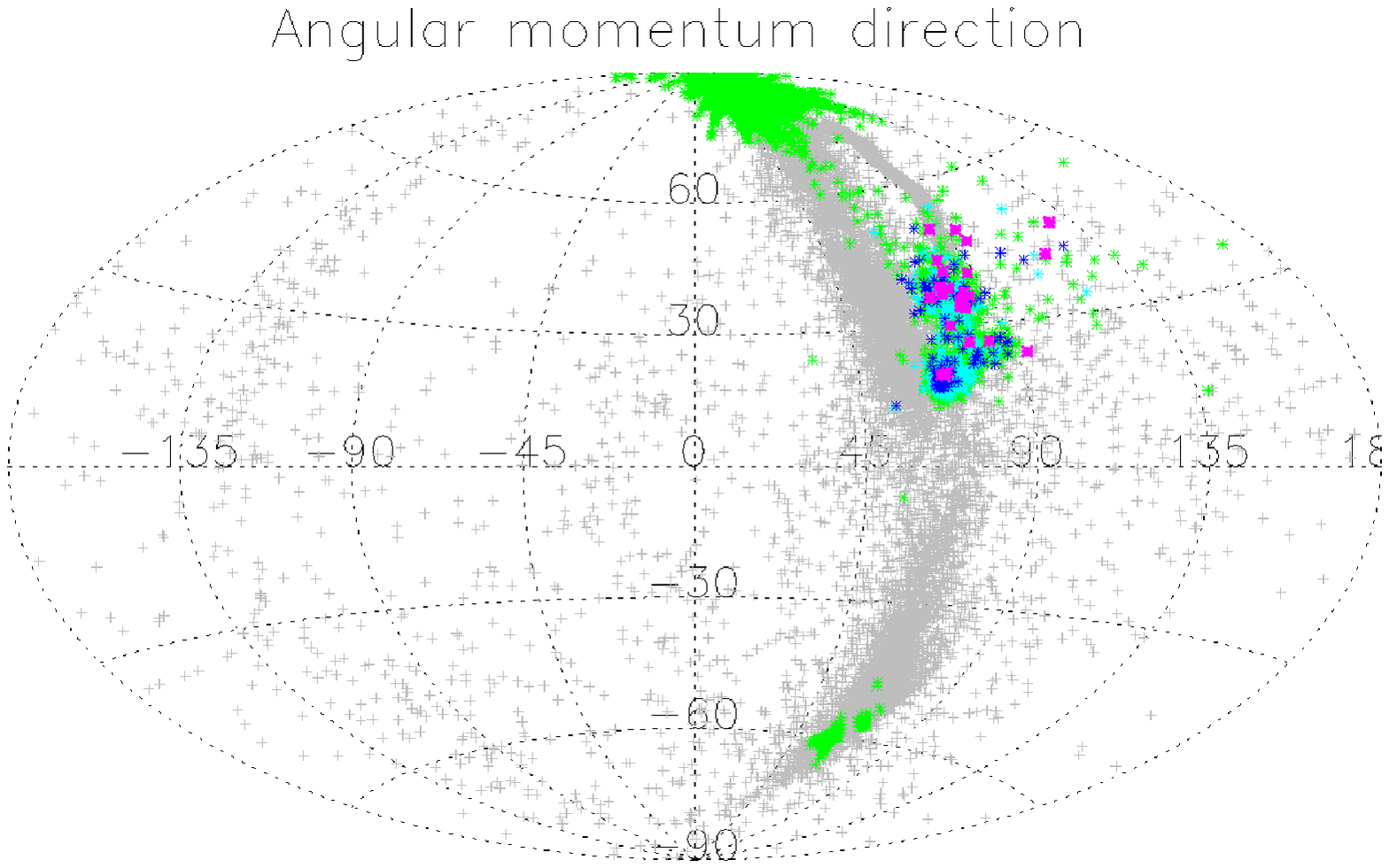,width=0.99\textwidth,angle=0}}
\end{minipage}
\caption{Angular momentum orientations for SPH particles (grey) and stars
(colours as in Fig \ref{fig:S1S2density}) in simulations S1 (left) and S2
(right) at $t=1955$.  Note that the inner stellar disc feature is rather small
in angular extent i.e. thin and planar in the left panel, but is geometrically
thick and significantly warped in the right.}
\label{fig:S1S2globus}
\end{figure*}

On the other hand, the parts of the clouds involved in the direct collision
experience a stronger shock in S2 than they do in S1 as the velocity vectors
in the former are almost directly opposing. Thus, despite the larger impact
parameter, the stronger angular momentum cancellation in S2 creates an inner
$r < 1$ disc as massive as that in S1. However, the ``mid-range'' disc largely
fails to form in S2.

\begin{figure*}
\begin{minipage}[b]{.48\textwidth}
\centerline{\psfig{file=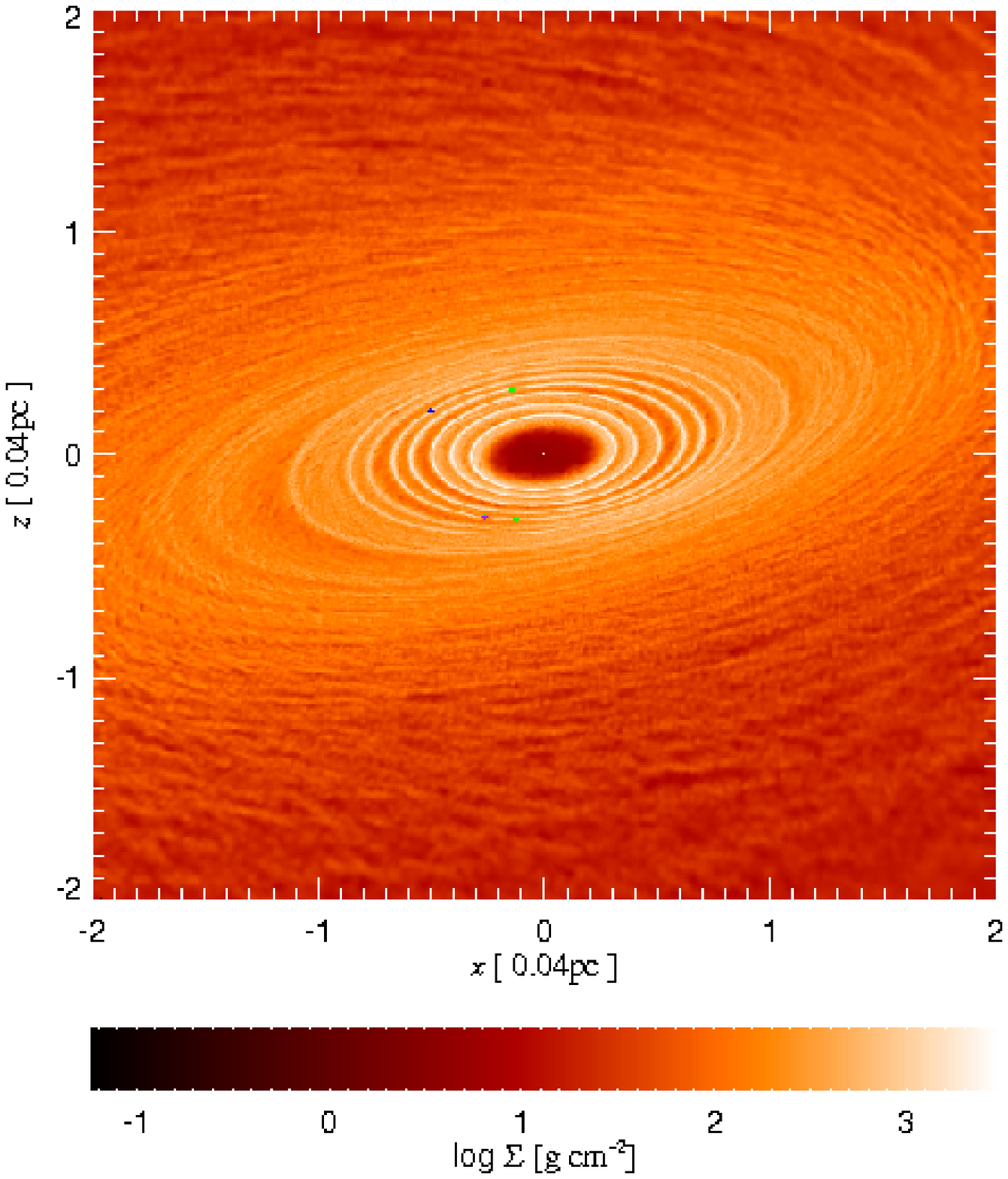,width=0.99\textwidth,angle=0}}
\end{minipage}
\begin{minipage}[b]{.48\textwidth}
\centerline{\psfig{file=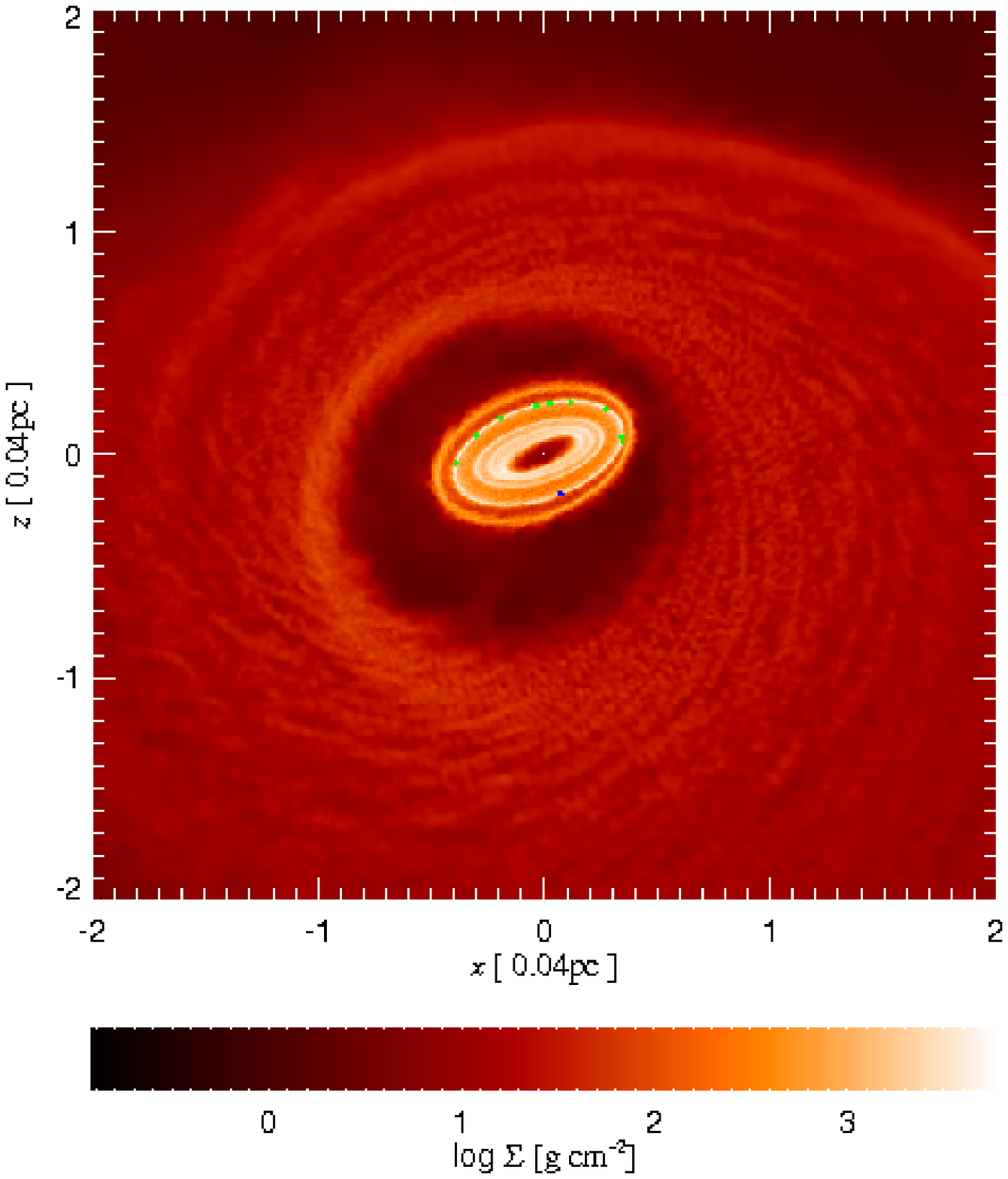,width=0.99\textwidth,angle=0}}
\end{minipage}
\caption{Projected gas density for the inner two arcseconds, showing
  stars forming (colours as in Fig \ref{fig:S1S2density}). {\bf Left:} Simulation S1 at $t=1000$; {\bf Right:} Simulation
  S2 at $t=400$. The times correspond to when first stars appear in these runs.}
\label{fig:S1S2densityarcsecond}
\end{figure*}

Another difference between S1 and S2 is the evolution and the structure of the
inner gaseous disc.  In both simulations the angular momentum of the gas infalling
onto the discs evolves with time, and hence the mid-planes of both discs undergo
a change of orientation in both $\theta$ and $\phi$ (we refer to this as `midplane rotation').
However the infalling gas seems to be more intermittent in S2 compared to S1, possessing a larger range of angular
momentum.  Whilst in S1 the accreting gas largely adds to the existing disc,
creating one large, relatively warped disc, in S2 the infalling gas causes
enhanced disc midplane rotation, and at later times ($t \sim 400$), actually
creates a second disc around the first at a different orientation. Over time,
the first (inner) disc aligns somewhat with the second one. This process
repeats at larger radii as gas continues to fall in with a different
orientation of angular momenta to that of the existing disc. The eventual result is
that the S2 disc is significantly more warped than its S1 counterpart. Figure
\ref{fig:S1S2densityarcsecond} compares the inner discs of S1 and S2 in the
process of forming.

\begin{figure}
\begin{minipage}[b]{.48\textwidth}
\centerline{\psfig{file=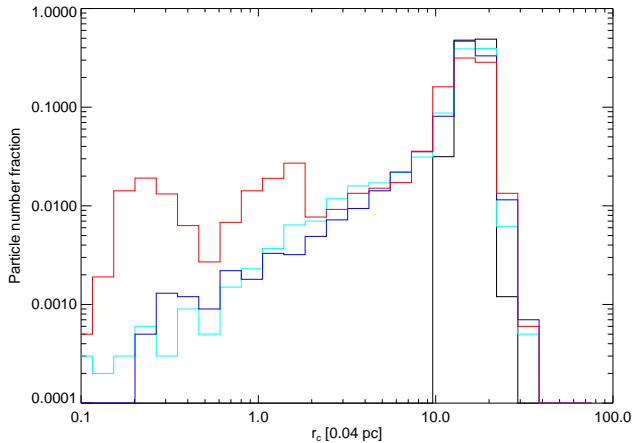,width=0.99\textwidth,angle=0}}
\end{minipage}
\caption{ The distribution of SPH particles over circularisation radius for
  simulation S2. The black, cyan, blue and red curves correspond to times $t=0,25,50,500$
  respectively.}
\label{fig:S2_rcirc_hist}
\end{figure}

The difference in gas distribution between the two simulations can be also
noted by looking at the respective histogram plots of circularisation radius
(Figure \ref{fig:S1_rcirc_hist} and Figure \ref{fig:S2_rcirc_hist}
respectively). A tail to small $r_{c}$ appears faster in S2, by $t = 25$
rather than $t = 50$ in S1. The radial profile of gas in S2 by $t = 500$ has a
gap between the inner and outer distributions, and is far more peaked than the
flatter, more even distribution of S1 at an equivalent time.
 
\subsubsection{Star formation}

Stars form earlier in S2 than in S1, and they form in the disc first, by $t =
400$. The preferential sites of star formation in the disc appear to be dense
concentric ring structures -- see Figure \ref{fig:S1S2densityarcsecond} -- we
discuss these in more detail in \S \ref{sec:rings}. Owing to the significantly
peaked distribution of gas (Figure \ref{fig:S2_rcirc_hist}) in S2, there is a
far greater amount of star formation at both small and large radii compared to
S1 at an equivalent time. This can be seen in Figure \ref{fig:S1S2density}. The
stars formed in the disc are again top-heavy and in the filaments are again
low-mass. The stars in the filaments show a tendency to form inside clusters
-- this effect is discussed in more detail in \S \ref{sec:S3}.

\subsubsection{Structures in angular momentum space}\label{sec:s2velocity}

Figure \ref{fig:S1S2globus} shows the orientation of the angular momentum
vectors for orbits of gas (in grey) and stars (in colour) in the Aitoff
projection, for S1 and S2 at the same times as in Figure
\ref{fig:S1S2density}. We identify stars of different masses using different
colours; green ($0.1 - 1 \msun$), cyan ($1 - 10 \msun$), blue ($> 10 \msun$),
and magenta ($> 150 \msun$).  As already noted, in S2 the secondary cloud
largely survives the collision and forms a stream, which later collapses into
stars. These stars can be found at the original orientation of that cloud in
the right panel in Figure \ref{fig:S1S2globus} (green at around $\theta =
-60^\circ$). The primary and secondary outer stellar populations are therefore
inclined at a large angle of $\theta \simeq 150^{\circ}$ to each other, as is
expected since they are the stellar remnants of the original clouds. The
angular momentum vector of the disc stellar population is oriented
at an angle in between that of the two outer stellar populations.

The other interesting feature of these plots is that the stellar disc  is
strongly warped. The origin of this warp is in the warping of the gaseous disc
rather than any secular evolution once the stars have formed.

\subsection{SIMULATIONS S3 \& S4}\label{sec:S3}

S3 and S4 are identical to S1 and S2 respectively, except for the value of the
cooling parameter $\beta$ which is set to a lower value of $0.3$. 

\subsubsection{Gas dynamics}\label{sec:S3dynamics}

\begin{figure*}
\begin{minipage}[b]{.48\textwidth}
\centerline{\psfig{file=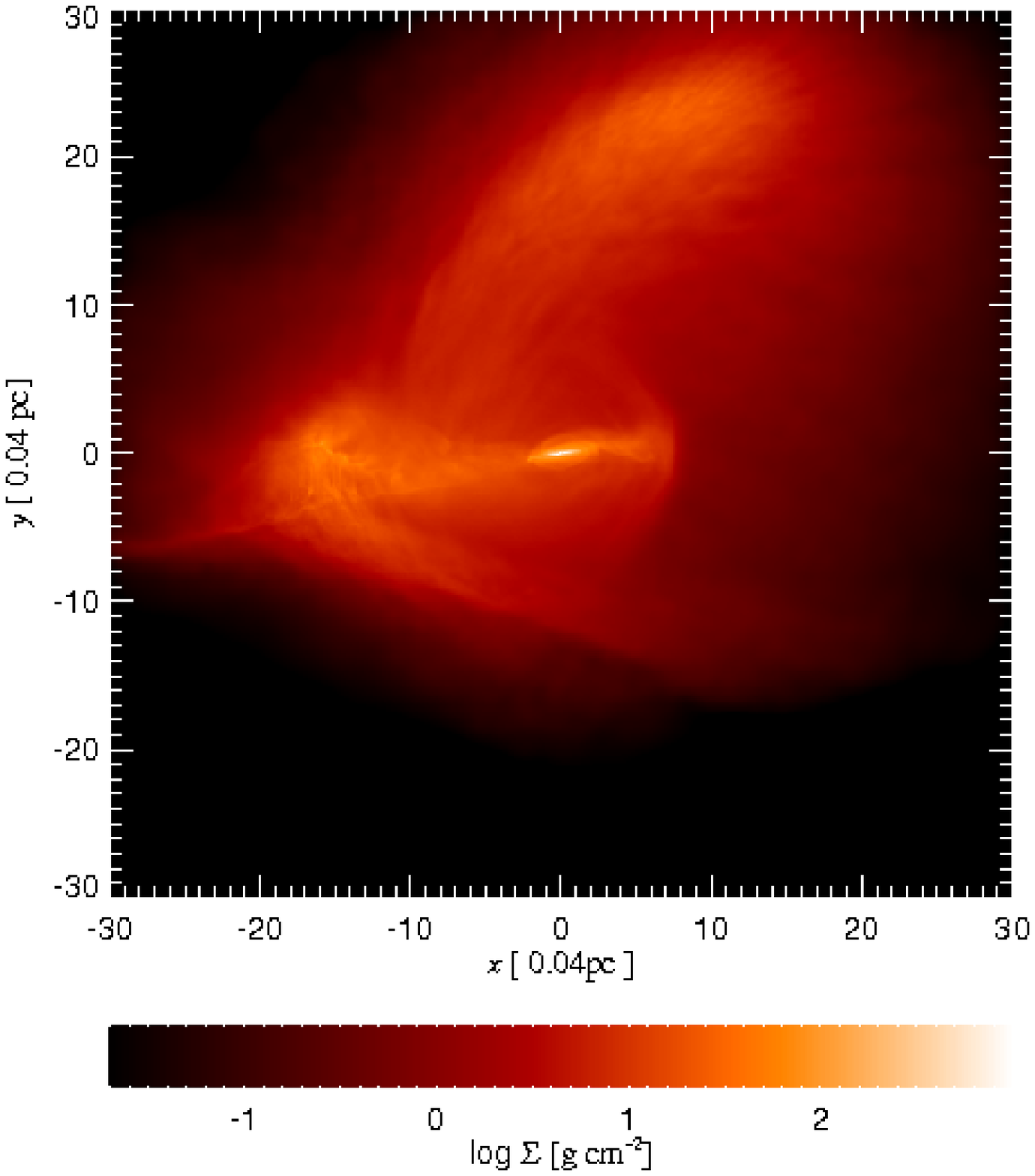,width=0.99\textwidth,angle=0}}
\end{minipage}
\begin{minipage}[b]{.48\textwidth}
\centerline{\psfig{file=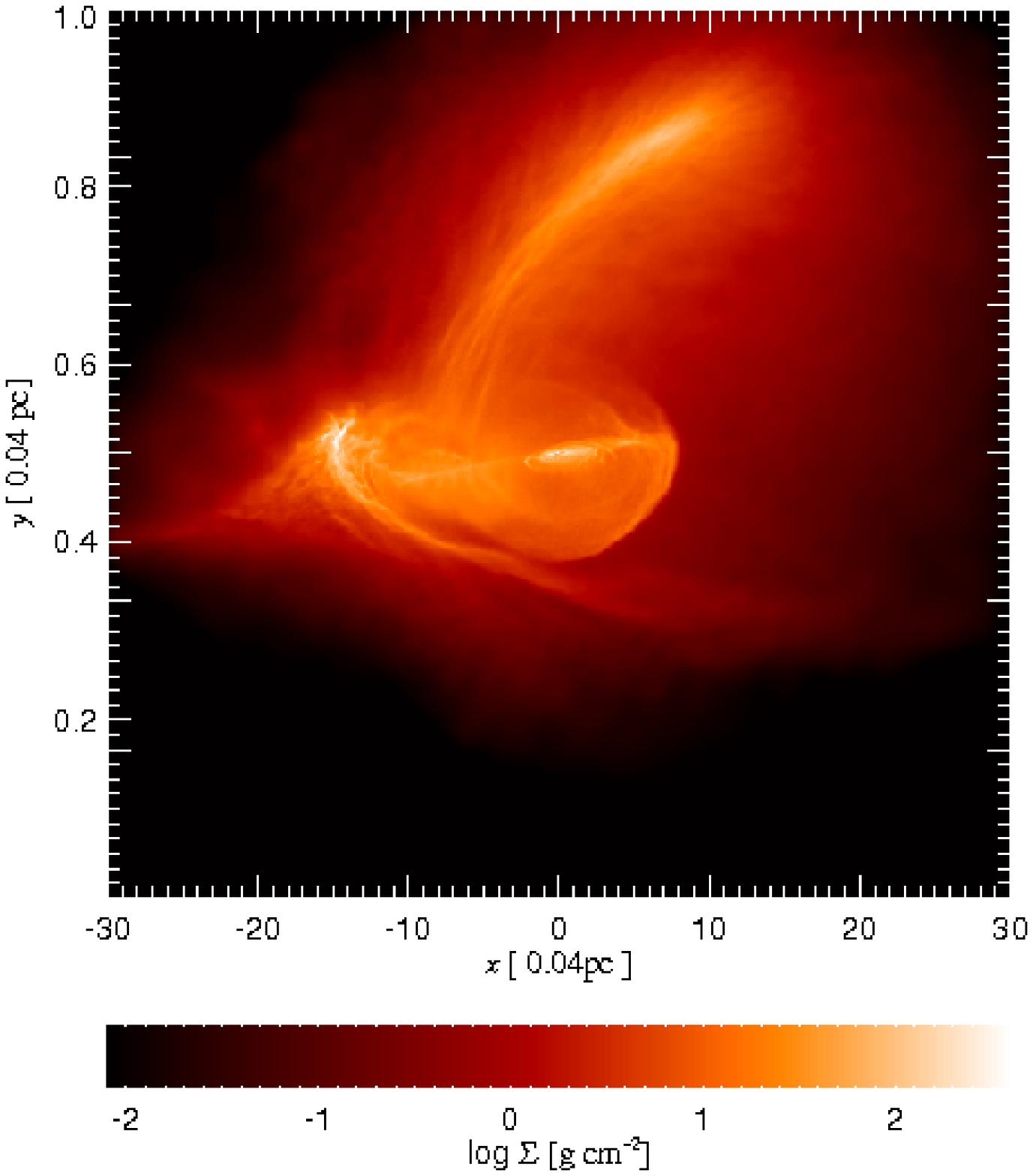,width=0.99\textwidth,angle=0}}
\end{minipage}
\caption{Comparison of projected gas density for simulation S1 (left panel,
  $\beta = 1$) and S3 (right panel, $\beta = 0.3$) at $t=150$. Note that
  filamentary structures are denser and better defined in S3 than they are in
  S1.}
\label{fig:S1S3densityglobus}
\end{figure*}

To show the effect of the faster cooling, Figure \ref{fig:S1S3densityglobus}
contrasts gas surface density plots of simulations S1 and S3 at time
$t=150$. The bulk dynamics of S3 are the same as in S1, but apart from the
innermost few arcseconds, maximum gas densities are higher in S3 than they are
in S1. This is natural as faster cooling gas can be compressed to higher
densities through shocks and gravity. For the same reason, the filaments are
much better defined in S3 than they are in S1. Similar conclusions can be
drawn from comparison of runs S2 and S4.

\subsubsection{Star formation}\label{sec:S3_SF}

Star formation occurs earlier in S3 than it does in S1. In particular, the first
stars appear at $t=150$ in S3 and only at $t\approx 1000$ in S1. This is
expected since faster cooling facilitates faster gravitational collapse
\citep[e.g.,][]{Gammie01,Rice05}.  The S2 to S4 comparison is less extreme but
still significant, with stars forming in S2 by $t = 400$ and in S4 by $t =
100$. In both $\beta = 0.3$ runs the initial sites of star formation were the
orbiting gas filaments, with the disc stellar populations forming slightly
later, at $t = 350$ (S3) and $t = 180$ (S4).

The clearly defined concentric ring structures of the S1 and S2 discs were
also seen in S3 and S4, although they did not provide the sites of star
formation in the latter cases, as fragmentation into sink particles occurred some
time before the rings began forming. In terms of the stellar mass spectra all
the stars formed in S3 and S4 are low mass ($\sim 0.1 - 1 \msun$), both
in the filament populations and in the disc. Again, this is a natural
consequence of a short cooling time \citep{Nayakshin06a,Shlosman89}.

\begin{figure}
\begin{minipage}[b]{.48\textwidth}
\centerline{\psfig{file=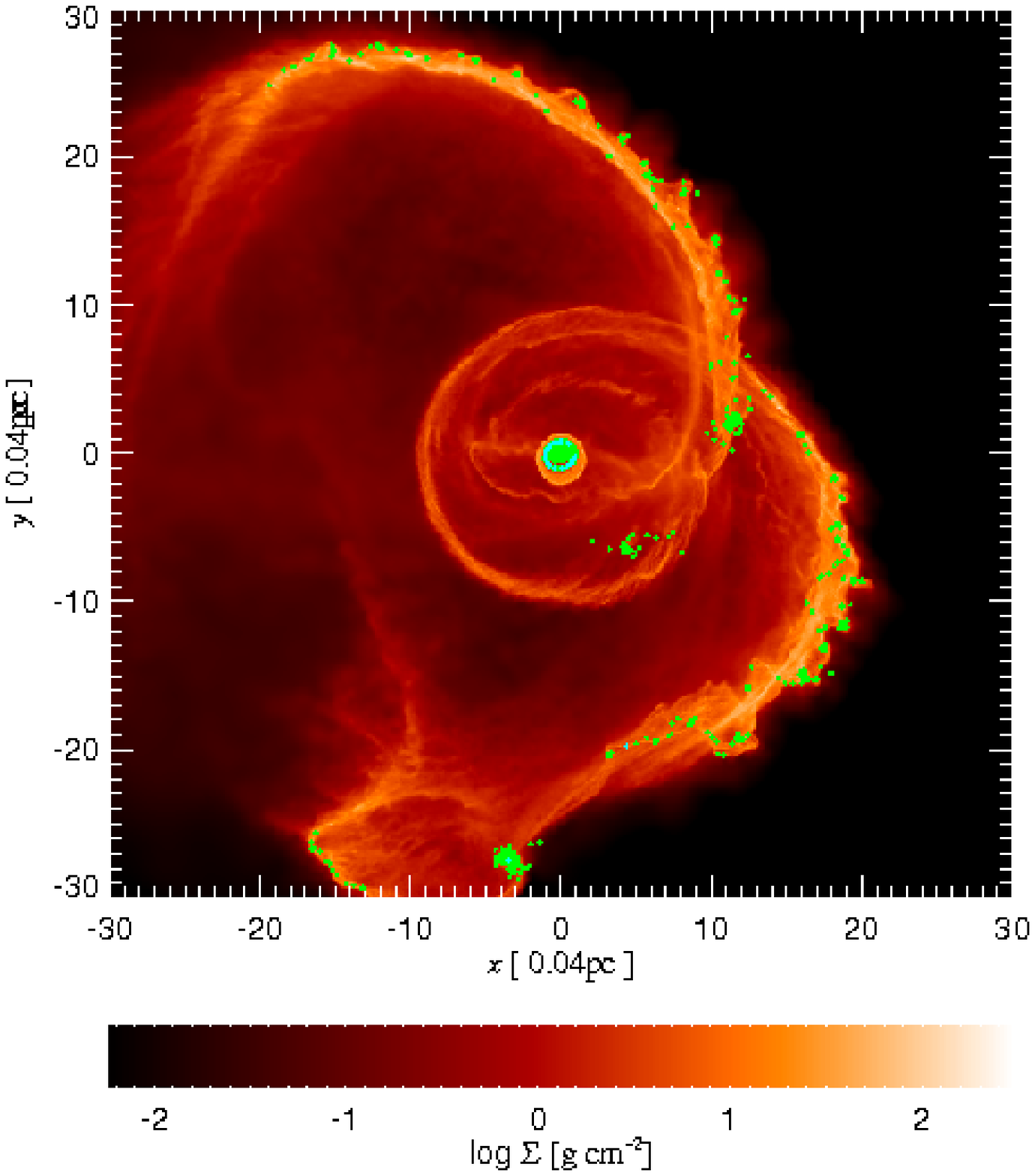,width=0.99\textwidth,angle=0}}
\end{minipage}
\begin{minipage}[b]{.48\textwidth}
\centerline{\psfig{file=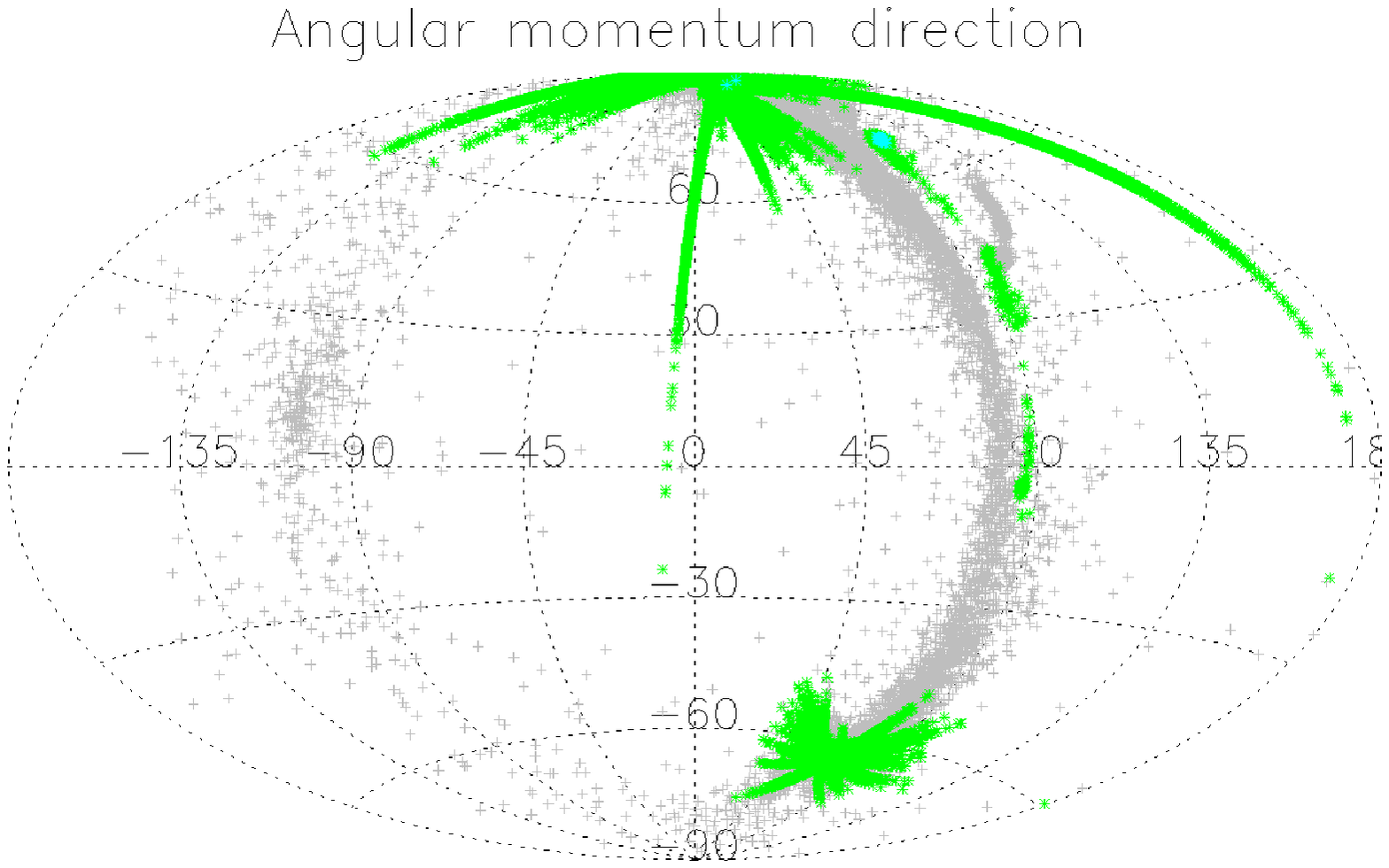,width=0.99\textwidth,angle=0}}
\end{minipage}
\caption{Projected gas density for simulation S4 (top) angular momentum of individual gas particles and stars (bottom) at
time $t=360$. In the angular momentum plot, gas is in grey whilst stars are colour-coded by mass, as they are in the gas density plot, as per \ref{fig:S1S2density}. The extended stellar structures in angular momentum space, such as the linear features seen, correspond here to extremely compact ``star clusters''. In the case of the most extended linear feature in the right-hand plot, velocity dispersion inside the cluster is comparable to its orbital velocity.}
\label{fig:S4densityglobus}
\end{figure}

\subsubsection{Clustered star formation}

The top panel of Figure \ref{fig:S4densityglobus} shows the surface density of
S4 some time after stars have begun to form. The bottom panel shows the orientation 
of the angular momentum of gas and star particles for the
same snapshot. In contrast to the equivalent plots for S1 and S2, one notices
extended linear structures, ``filaments'' or ``arms'' in Figure
\ref{fig:S4densityglobus}.  These structures, extended in angular momentum
space, are in fact produced by the most compact features in coordinate space
-- the star clusters.  These clusters are seen visually in the top panel of
Figure \ref{fig:S4densityglobus}. The velocity dispersion of the stars in these very dense clusters
is high; some are comparable to the orbital velocities of the clusters
themselves. For example, in the bottom panel of Figure
\ref{fig:S4densityglobus}, the two longest continuous ``filaments'', separated
by 180$^\circ$ in $\phi$, are produced by the densest and most massive cluster
located at $z\approx 0$, $x = 4$, $y\approx -28$ in the top panel. Due
to its location close to the $y$-axis, the $y$-component of the angular
momentum of any star that belongs to the cluster, $l_y = -x v_z + z v_x
\approx 0$. Therefore the cluster's stars form an incomplete plane $l_y
\approx 0$ in the angular momentum space, producing the two (actually one)
very long ``filaments'' in the bottom panel of Figure
\ref{fig:S4densityglobus}. This cluster is a remnant of cloud 1, and its mass
is $\sim 3000$ solar masses.

In fact most of the stars in simulations S3 and S4 belong to a cluster. This
clustered mode of star formation in the fast-cooling runs is due to the gas
collapsing promptly and thus forming very massive dense gas halos. We suspect
that these results would have been different had we been able to treat
radiative feedback from star formation in our simulations.  We note that in
such a dense environment, any feedback from star formation activity, such as
radiation or outflows, would not have been able to escape easily and would
therefore have heated the surrounding gas, suppressing further fragmentation
and increasing the Jeans mass \citep{Nayakshin06a,KrumholzEtal07}. Rather than
the multiple low-mass stars in a cluster we would perhaps have seen fewer but
higher mass stars. The mass spectrum of stars would probably also change if
the colliding clouds had net rotation (spin) or a turbulent structure before
the collision, providing some stability against collapse. All these effects are
however minor as far as orbital motion of the stars is concerned, which is the focus of our paper.

\subsubsection{Inner fluctuating disc}\label{sec:S4_inner}

\begin{figure*}
\begin{minipage}[b]{.48\textwidth}
\centerline{\psfig{file=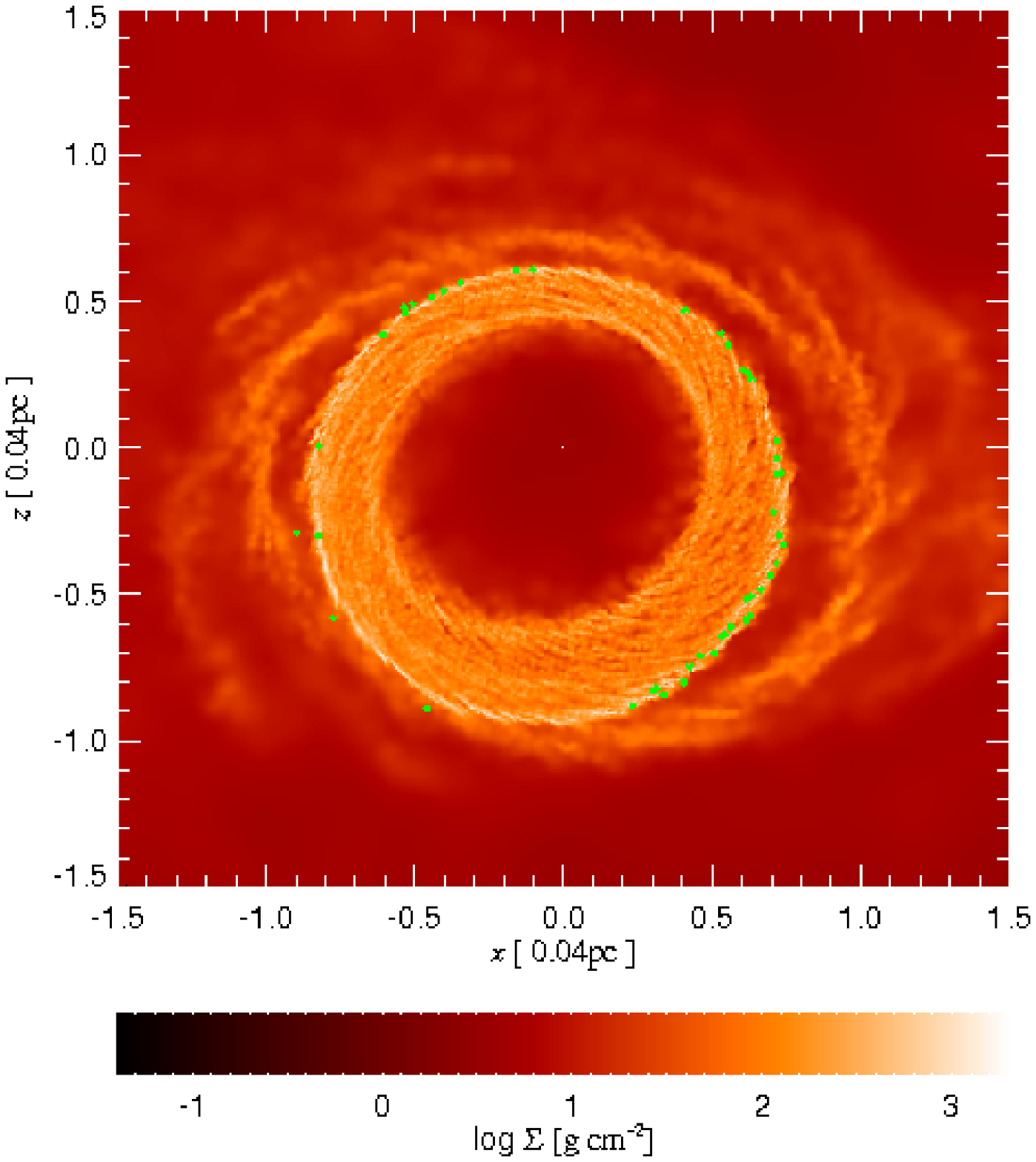,width=0.99\textwidth,angle=0}}
\end{minipage}
\begin{minipage}[b]{.48\textwidth}
\centerline{\psfig{file=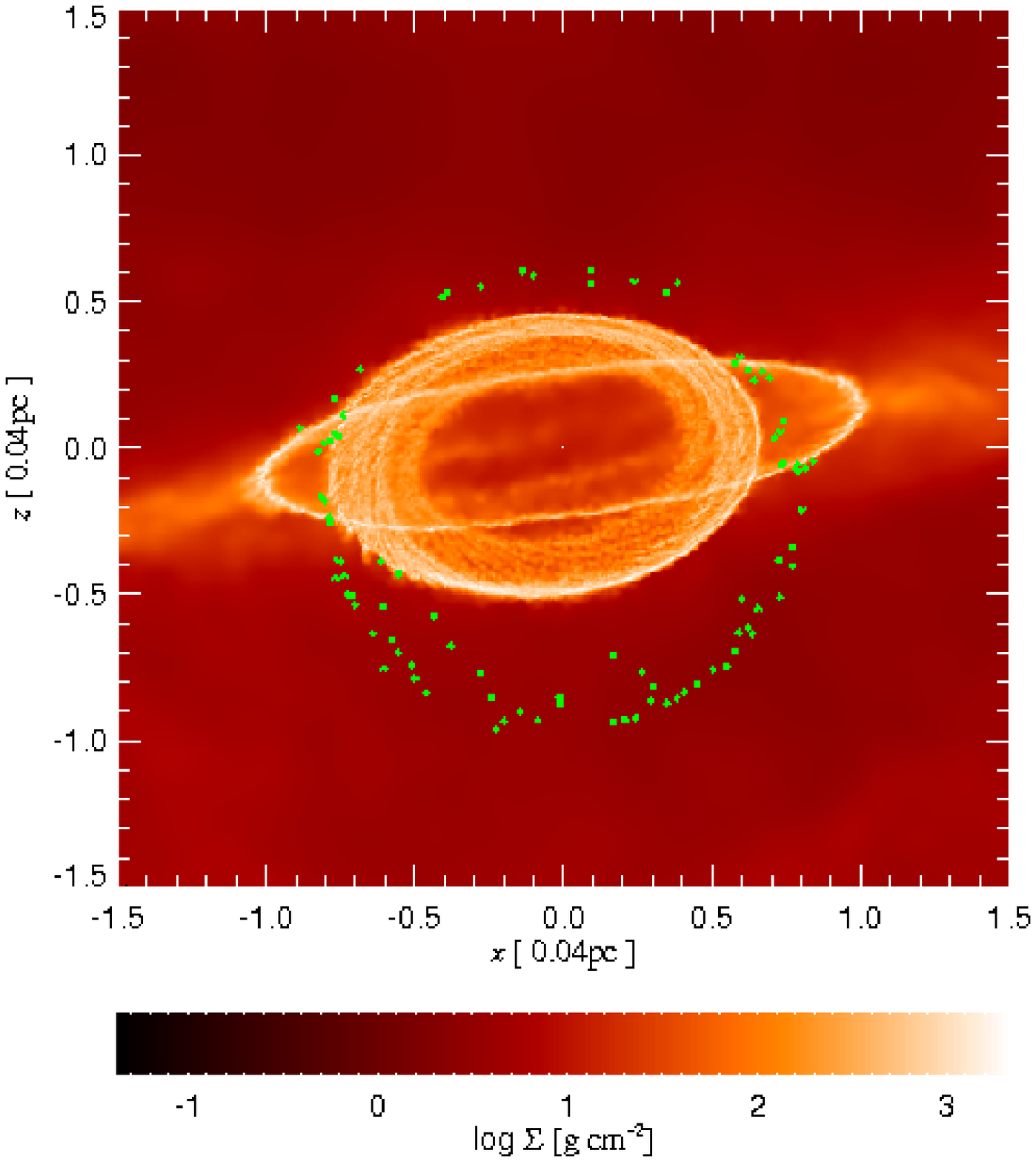,width=0.99\textwidth,angle=0}}
\end{minipage}
\caption{The inner 1.5 arcseconds of the simulation S4 at $t=180$ (left
  panel) and $t=260$ (right panel). Note that stars born in the gaseous disc at
  time earlier than $t=180$ kept their orbital orientation whereas the gaseous
  disc has evolved due to deposition of new matter from larger radii.}
\label{fig:S4_inner_disk}
\end{figure*}

Simulation S4 is the fast-cooling equivalent of S2. As such, the inner gaseous
disc in S4 is subject to a similar intermittent infall of gas from larger
radii. The newly arriving gas tweaks the disc orientation significantly,
causing it to undergo midplane rotation (refer back to \S
\ref{sec:s2dynamics}).  The difference here however is that disc is
fragmenting into stars whilst its orientation is changing. Interestingly,
these stars do not follow the evolving disc orientation but rather remain in
their original configuration.  This is clearly seen in Figure
\ref{fig:S4_inner_disk}, where the stellar disc remembers the ``old''
orientation of the gaseous disc in which the stars were born, whereas the
gaseous disc evolves to quite a different orientation. The possibility of this
effect taking place was suggested by \cite{NC05}. These authors found that
stars will not follow disc midplane changes if these occur faster than the
``critical rotation time'', which is estimated to be about 500 code
units for the disc in Figure \ref{fig:S4_inner_disk}. Our simulations are thus
consistent with these predictions as the disc orientation changed on a
timescale of just $80$ code units.

As a result of a several star forming events in gaseous discs of different
orientations, the inner stellar disc in S4 is very much different from that
found in S2, and even more so when compared to S1. In the latter
cases, when cooling is more gradual (i.e. $\beta=1$), stars only form in the inner gaseous disc once it 
has settled into a relatively stable orientation. Hence the resulting distribution of stars is
correspondingly thin (although in S2 it is significantly thicker than S1, cf. Figure \ref{fig:S1S2globus}). Such a distribution is broadly consistent with the observed orbits of young massive stars in the
clockwise disc in the GC \citep{PaumardEtal06}. Due to gaseous disc midplane changes
coupled with quicker fragmentation, then, the stellar disc in S4 is much
thicker, with $H/R\sim 1$. This particular simulation thus fails to account for the most
prominent feature of the observational data; namely, a thin inner stellar disc.

\subsection{SIMULATIONS S5 \& S6}\label{sec:S5}

\begin{figure}
\centerline{\psfig{file=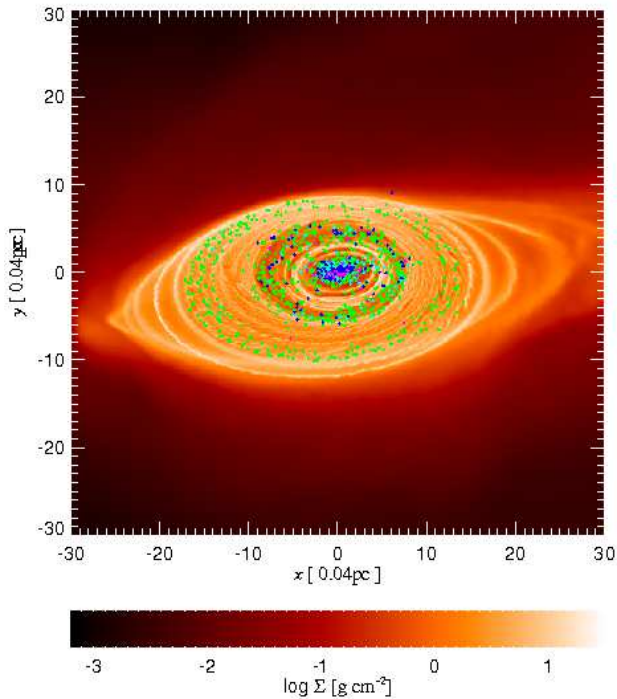,width=0.5\textwidth,angle=0}}
\caption{Projected gas density for S5 at $t=2696$, showing stars as well
  (colours as in Fig \ref{fig:S1S2density}). Note that the stellar populations
  all possess similar values for eccentricities.}
\label{fig:S5}
\end{figure}

Simulation S5 is somewhat distinct from all the rest due to its small impact parameter,
$b$, between the gas clouds (see Table 1). As a result of this fact and the
relatively large value of $\beta$, a higher degree of mixing is
achieved. Figure \ref{fig:S5} shows a snapshot from this simulation, showing
both gas and stars. Compared with all the other runs, the accretion
disc extends out to as much as $r \sim 20$ and represents a very coherent disc
with little warping. Outer gas filaments do not form
due to an almost complete mixing of the original gas clouds.

This simulation is an excellent example of how just a small change in initial conditions
can yield drastically different results. Comparing the parameters for S5 with those for S2 (refer
to Table \ref{table1}), we see that there are only relatively minor differences (the largest one being
the different in impact parameter). However, towards later times the behaviour of the gas diverges
considerably, with S5 ending up with a single, extended disc and similarly-oriented stellar
populations and S2 possessing a small, highly warped stellar disc together with outer stellar populations
inclined at a large angle from each other.

Simulation S6 is again different in that it possesses
the largest collision velocity of all the simulations, whilst retaining a
cooling parameter of $\beta = 1$. One therefore expects the resulting thermal
expansion of the clouds to significantly modify the velocities of the
particles and encourage mixing. However, since the impact parameter is
relatively large, this simulation actually has more in common with S2 than
with S5, particularly in the inner parts where the gas forms into a double
disc structure (refer back to \S \ref{sec:s2dynamics}). The gas dynamics are therefore
largely equivalent to S2, and as such we will not discuss this simulation in detail.

\subsection{Star formation in rings}\label{sec:rings}

One particular interesting feature of our simulations is that the disc surface
density exhibits slightly eccentric ``rings'' in the inner $0.5-1''$. Star
formation occurs preferentially in these rings (see Figure
\ref{fig:S1S2densityarcsecond}) in the slower cooling runs. The inner discs in
runs S1-S6 are eccentric, warped, with a radial gradient in both of these properties,
and evolving under the influence of a continuous but variable infall of
material from larger radii, in a fixed (BH + stellar cusp) potential. It is
difficult to disentangle these various effects from one another to understand
the origin of the rings in detail, but we believe that they are partly a consequence
of a fast cooling time (since $t_{cool} \propto \beta R^{3/2}$) coupled
with most likely a number of the above properties of the gas in the inner
parts. We base this conclusion on our unsuccessful attempts to reproduce the
rings in cleaner purpose-designed simulations. In particular we ran
simulations with eccentric and circular, flat or warped discs, but found no
ring structures. Similarly, \cite{NayakshinEtal07} did not encounter any such
structures in their simulations of circular or eccentric self-gravitating
discs. We plan to continue to investigate the cause of the rings, and in
particular to determine whether they are a numerical or physical artifact.  It
should be noted at this point that the effect of the rings on our model and
overall conclusions is (fortunately) negligible, as they only affect star
formation within the inner $\sim 0.5''$, a region in which fragmentation would
be suppressed through accretion luminosity \citep{Nayakshin06a}, if we were to
include this in our model.

\section{Radial distribution of stars}\label{sec:radial}

\begin{figure*}
\begin{minipage}[b]{.48\textwidth}
\centerline{\psfig{file=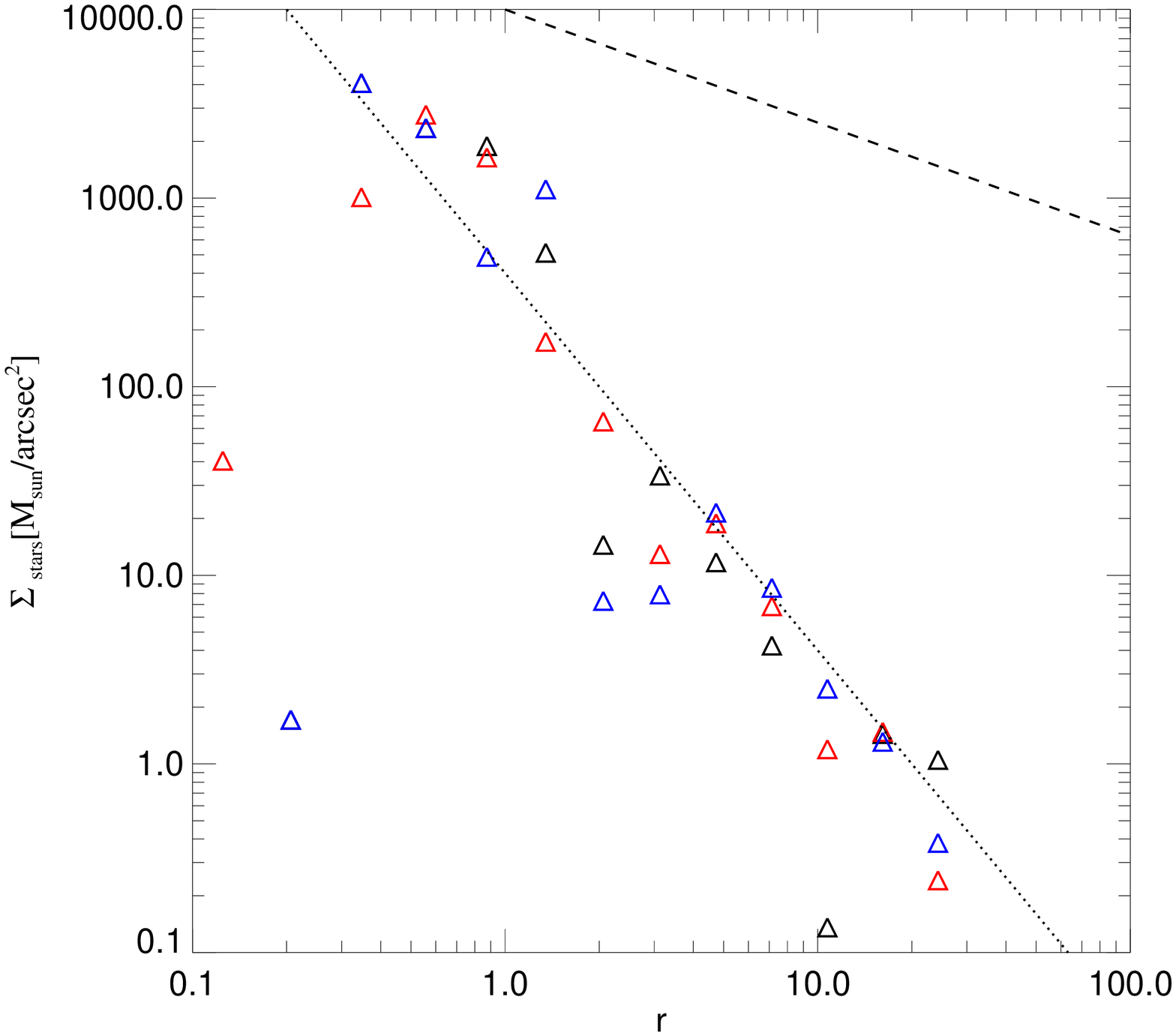,width=0.99\textwidth,angle=0}}
\end{minipage}
\begin{minipage}[b]{.48\textwidth}
\centerline{\psfig{file=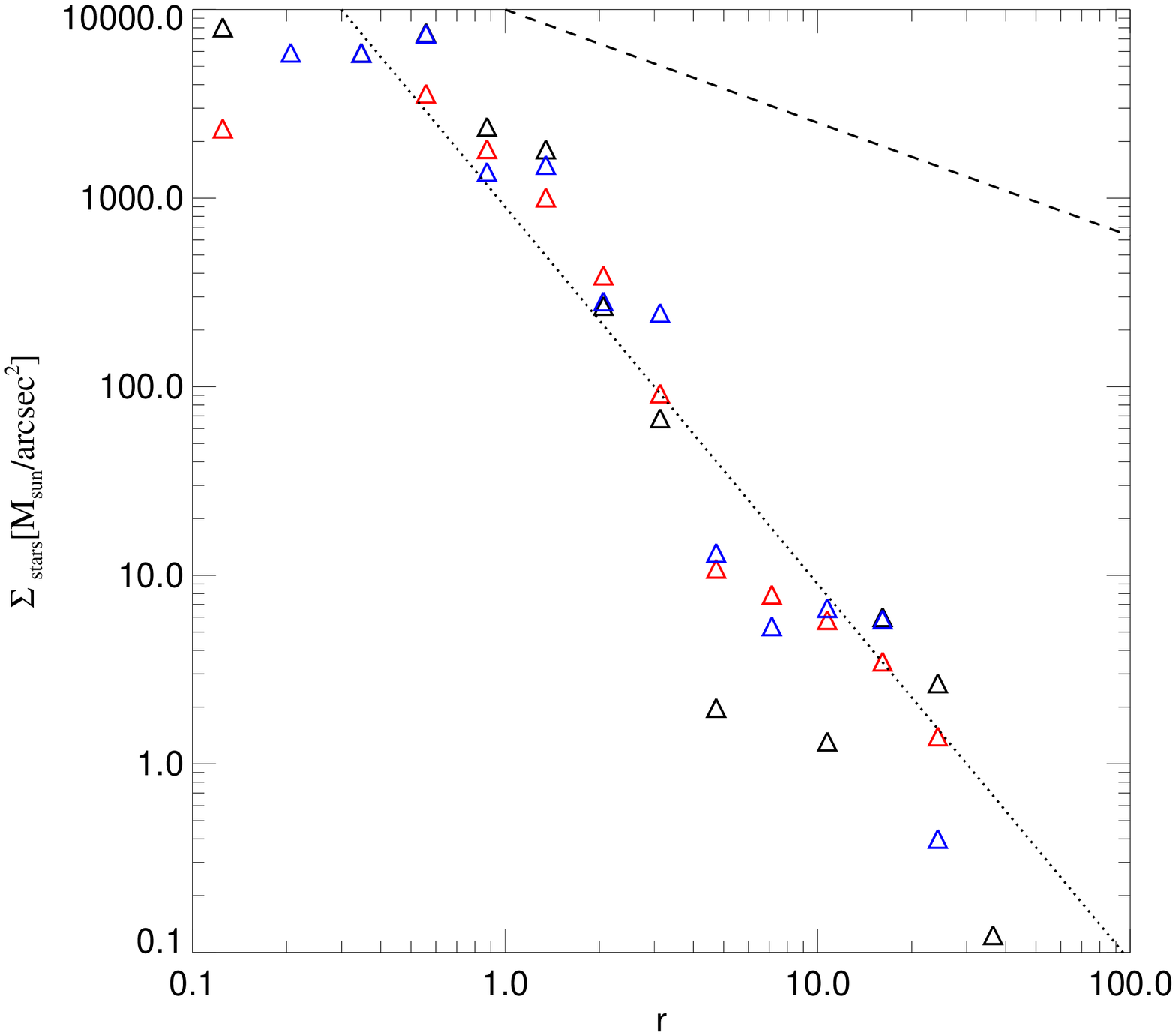,width=0.99\textwidth,angle=0}}
\end{minipage}
\begin{minipage}[b]{.48\textwidth}
\centerline{\psfig{file=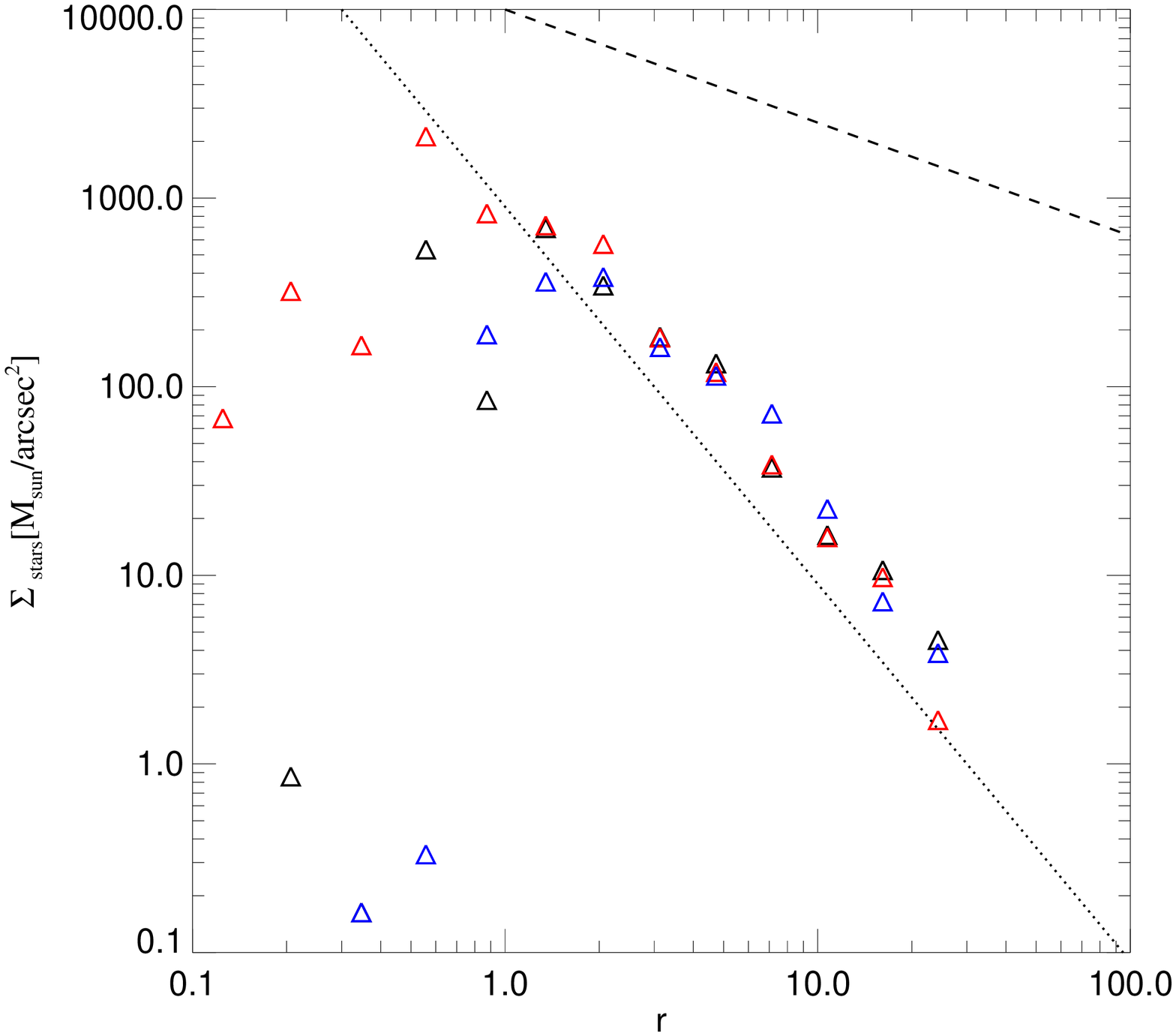,width=0.99\textwidth,angle=0}}
\end{minipage}
\begin{minipage}[b]{.48\textwidth}
\centerline{\psfig{file=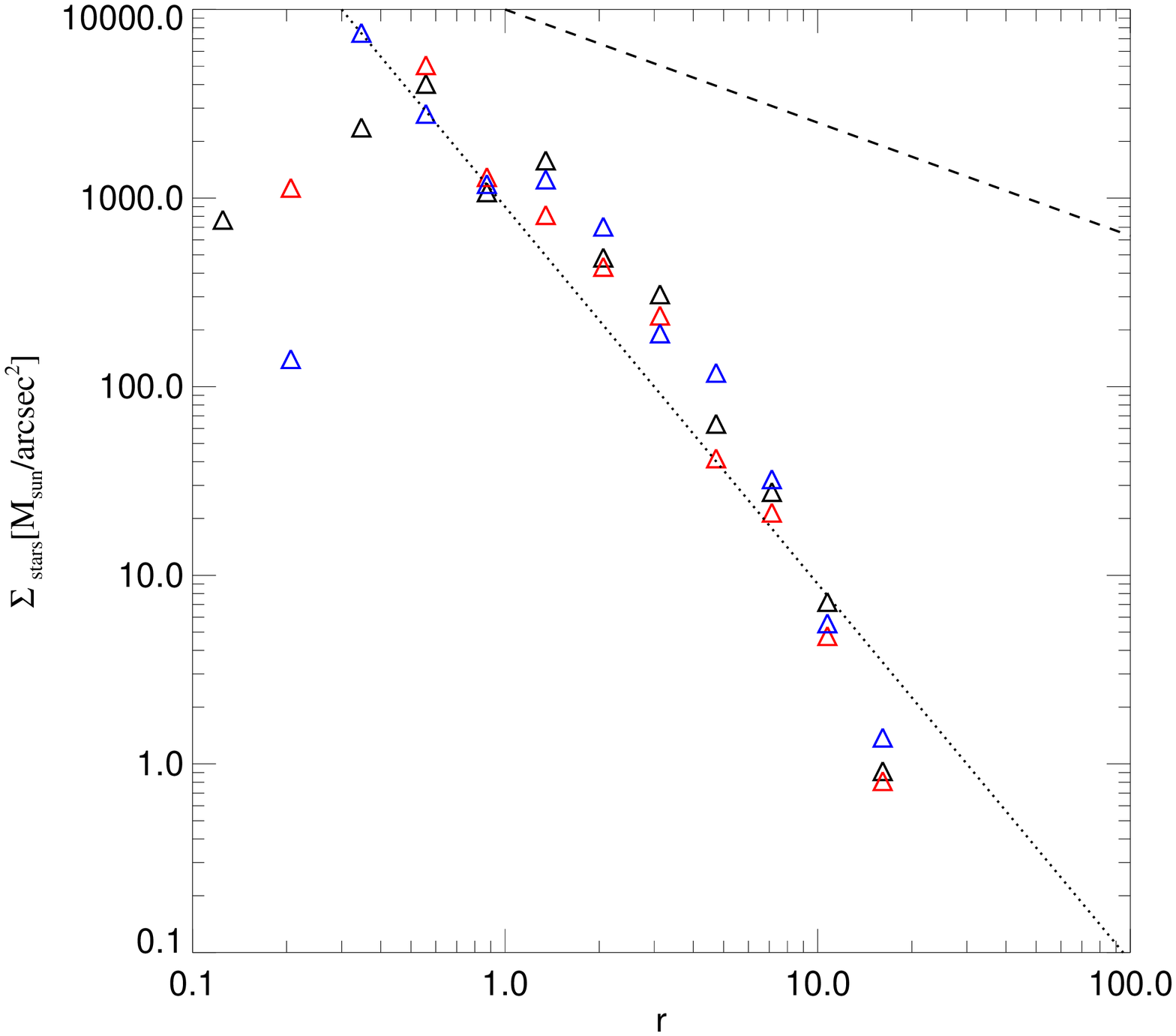,width=0.99\textwidth,angle=0}}
\end{minipage}
\caption{Stellar surface density, $\Sigma_*$,  versus projected radius, for runs S1 (top left, at $t=1955$),
  S2 (top right, at $t=2243$), S3 (bottom left, at $t=1396$) and S5 (bottom right, at $t=2696$). 
  Different colours show the three different orientations, along $z$, $x$ and $y$ axes (black, red, and
  blue respectively). Power laws are overlaid for $r^{-2}$ (dotted) and $r^{-3/5}$ (dashed). All four plots 
  correspond to the maximum time to which each simulation has been run.}
\label{fig:sigma_vs_r}
\end{figure*}

Interestingly, the most robust result from all of the 6 runs completed is the
fact that distribution of stellar mass in an annulus, $dM_* = \Sigma_*(R) 2\pi
R dR$, versus radius roughly follows the law
\begin{equation}
\Sigma_*(R) \propto \frac{1}{R^2}\;.
\label{sigma_vs_r}
\end{equation}
We show these distributions for runs S1, S2, S3 and S5 (all the simulations that have
progressed far enough to transform a significant fraction of the gas into stars). To
facilitate comparison with observational data, we use {\em projected} radius 
instead of the proper 3D radius, which is not directly known in
observations of young stars in the GC \citep[e.g.,][]{GenzelEtal03}. As the
viewing angle of the stellar system modelled here is arbitrary, we chose to
plot the distributions along the three axes of the simulations, so that $r =
\sqrt{x^2 + y^2}$ for the black symbols, $r = \sqrt{y^2 + z^2}$ for the red
symbols, and $r = \sqrt{x^2 + z^2}$ for the blue symbols. The lines in the
Figure show $r^{-2}$ (dotted) and $r^{-3/5}$ (dashed) power laws. The latter
corresponds to that predicted by {\em non self-gravitating} standard
accretion disc theory.

This result is remarkable given very large differences in the 3D arrangments
of the stellar structures in our simulations. The $1/R^2$ law is in good agreement
with the observed distribution of ``disc'' stars
\citep{PaumardEtal06}. We have not been able to attribute a simple anaytical explanation to 
this scaling, other than to suggest that cancellations of angular momenta in shocks
appears to be very efficient in driving gas to smaller radii. On the referee's request however
we have looked at the distribution of the gas for a given N-body velocity profile emerging from the initial
cloud collision, and have attempted to fit the resulting circularisation radius $r_{c}$ distribution to
our result for $\Sigma_*(R)$. The thermal expansion of our clouds as they collide can be modelled to a
first approximation as an isotropic, delta-function velocity profile, which for small values of $r_{c}$ yields a slope of $\Sigma \propto R^{-1}$.
Introducing a gaussian velocity profile into the isotropic distribution, centered about low velocities, acts to steepen this slope closer to the desired $R^{-2}$.
Beyond this simple N-body model, shocks and small-scale collisions
elsewhere in the simulation will drive gas deeper into the potential well, increasing the slope even further. It would be interesting to
explore a larger parameter space to quantify the robustness of these results.

\section{Sgr A* feeding during and after the cloud
  collision}\label{sec:accretion}

\begin{figure*}
\begin{minipage}[b]{.48\textwidth}
\centerline{\psfig{file=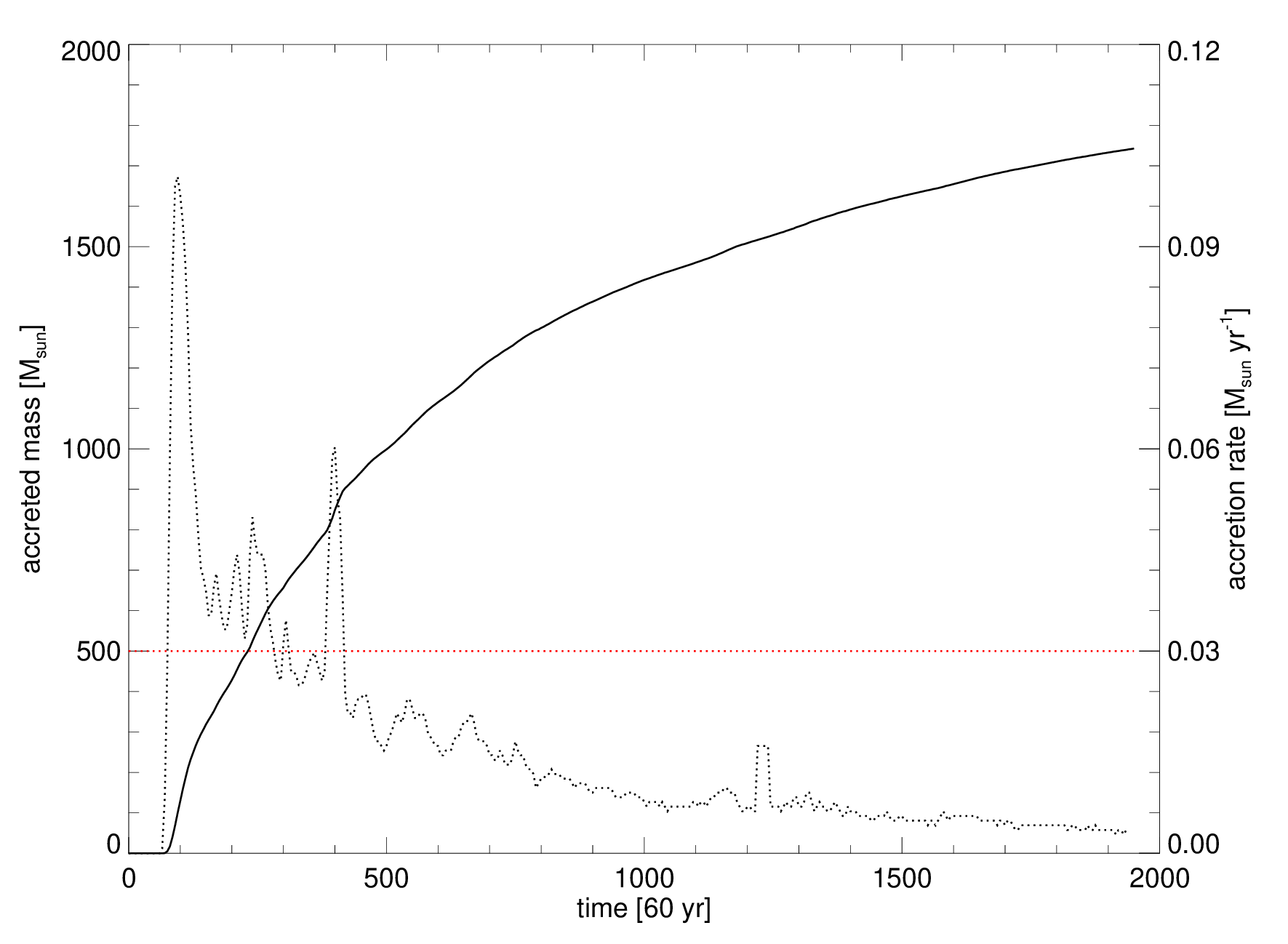,width=0.99\textwidth,angle=0}}
\end{minipage}
\begin{minipage}[b]{.48\textwidth}
\centerline{\psfig{file=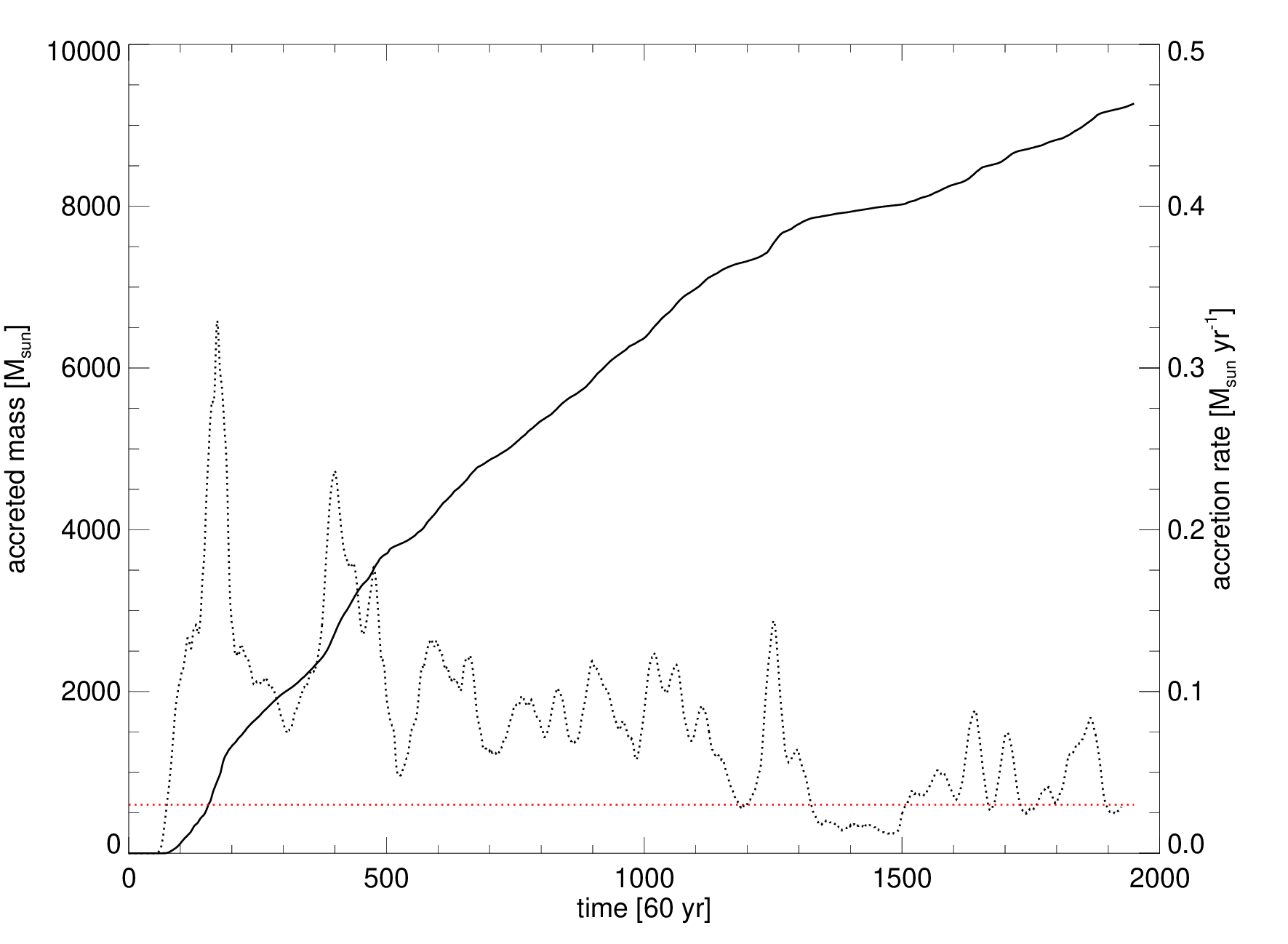,width=0.99\textwidth,angle=0}}
\end{minipage}
\begin{minipage}[b]{.48\textwidth}
\centerline{\psfig{file=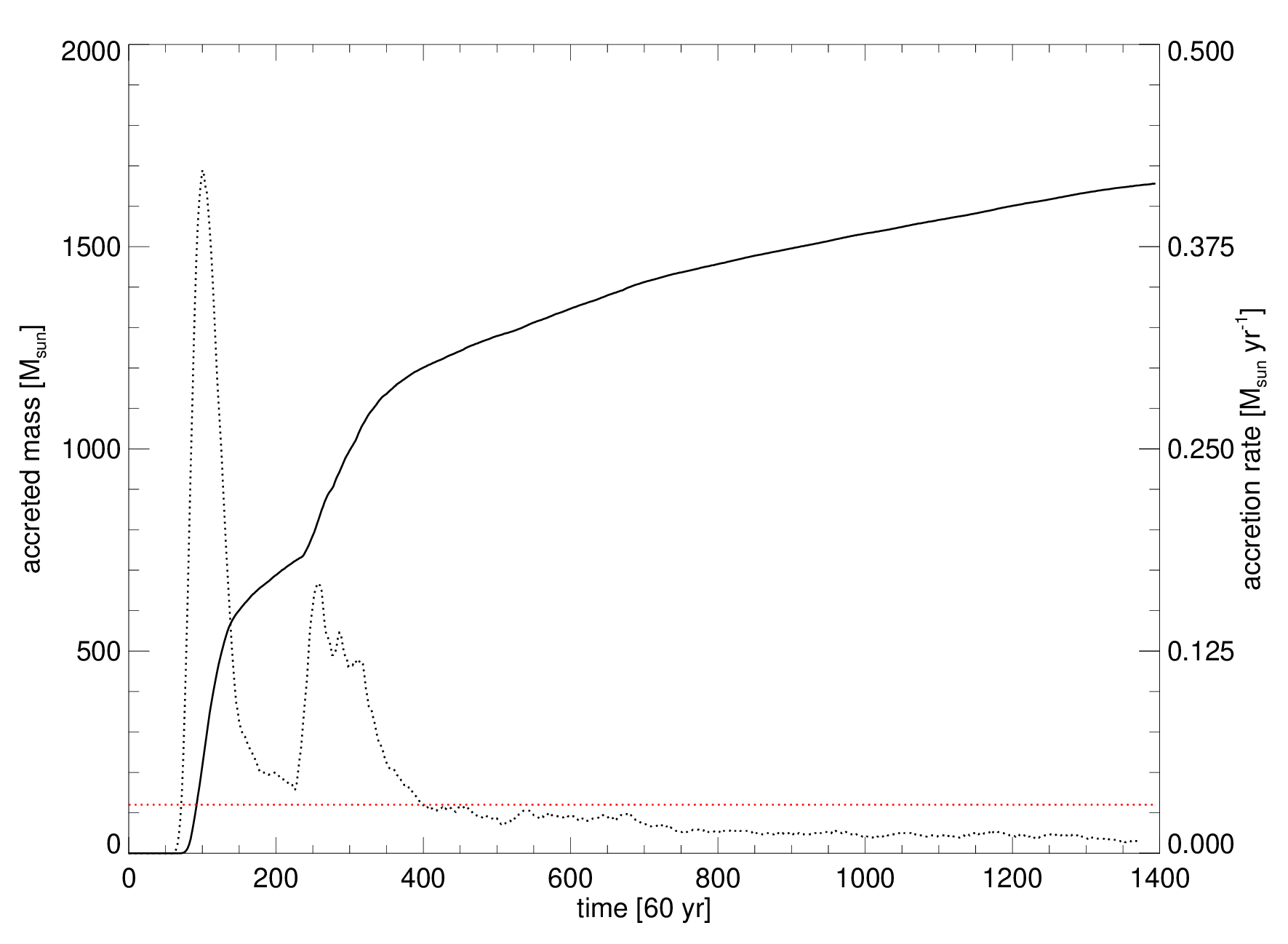,width=0.99\textwidth,angle=0}}
\end{minipage}
\begin{minipage}[b]{.48\textwidth}
\centerline{\psfig{file=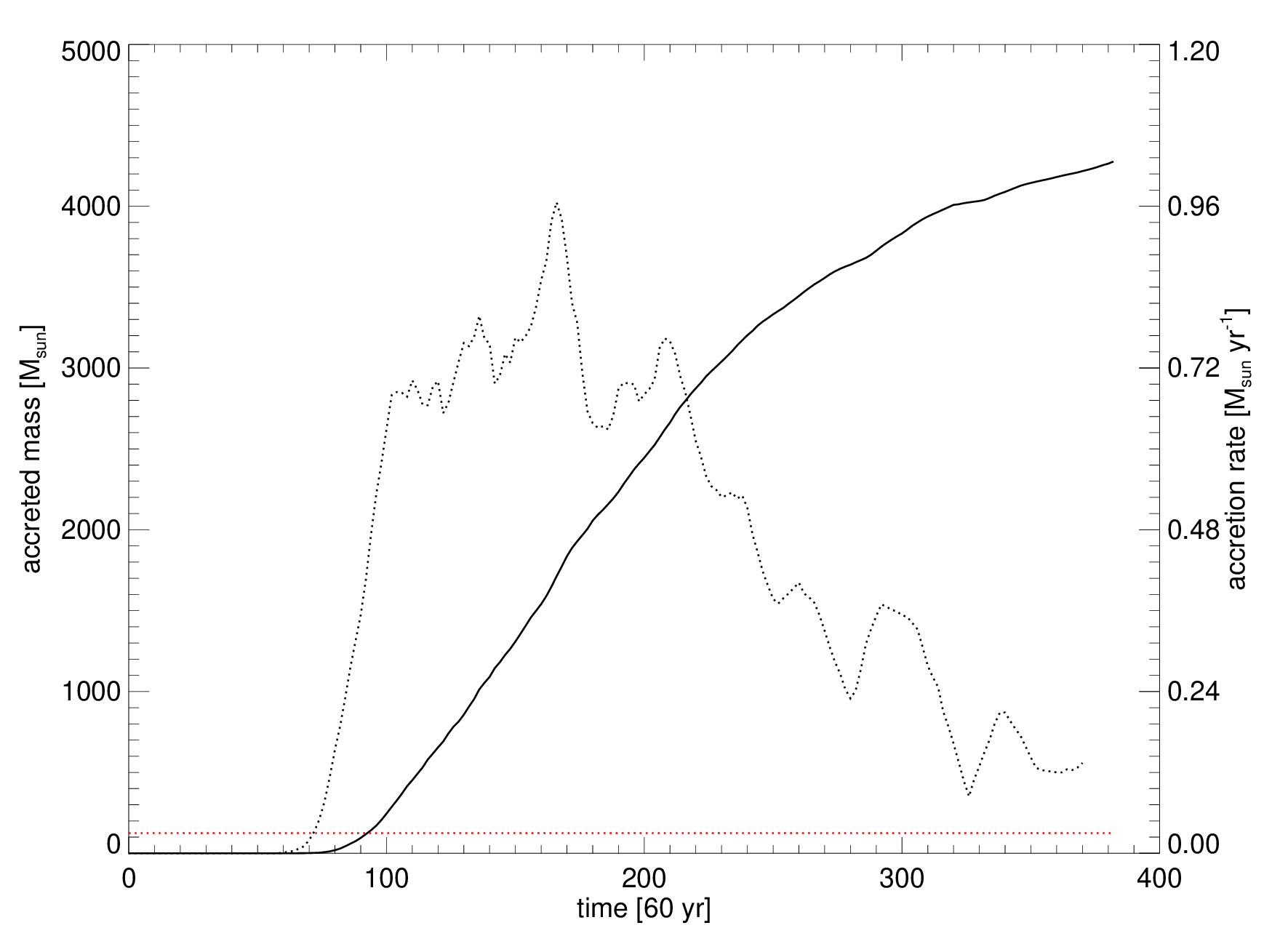,width=0.99\textwidth,angle=0}}
\end{minipage}
\caption{The time evolution of the accreted mass onto the central black hole (solid line)
and the corresponding accretion rate (dotted line) compared to the Eddington limit for
Sgr A* (dotted red line), for simulations S1 (top left), S2 (top right), S3 (bottom left),
and S4 (bottom right)}
\label{fig:accretionhistory}
\end{figure*}

Figure \ref{fig:accretionhistory} shows the mass accretion rate for the
central black hole for simulations S1-S4. Formally, the Figure indicates that
our simulations provide a sustained super-Eddington accretion rate (the Eddington
accretion rate for Sgr A* being $\sim 0.03 \msun$ yr$^{-1}$), in some cases for between $10^{4}$ to
10$^5$ yrs. However it must be remembered that we do not resolve gas dynamics
inside the accretion radius. We expect that material will form a disc there,
and accretion will proceed viscously. The viscous time scale, $t_{\rm visc}$,
depends on the temperature in the disc midplane and the viscosity parameter,
$\alpha$ \citep{Shakura73}. The midplane temperature in the inner arcsecond is
around $10^3 K$ in both the standard \citep[non self-gravitating,
e.g.,][]{NC05} and the self-gravitating regimes \citep{Nayakshin06a}, yielding
$H/R \sim 0.01$.  Hence the viscous time is
 
\begin{equation}
t_{\rm visc} = 6\times 10^6 \;\hbox{years}\; \alpha_{0.1}^{-1} \left[\frac{R}{100
H}\right]^2\; R^{3/2}\;,
\end{equation}

\noindent where $\alpha_{0.1} = \alpha/0.1$ and $R$ is the radius of the disc in arcseconds.
With these fiducial numbers, the viscous timescale coincides with the age of
the young massive stars in the GC \citep[e.g.,][]{KrabbeEtal95,PaumardEtal06}. As
$\alpha$ is highly uncertain, we shall then consider the two opposite cases.
If $\alpha = 0.1$, $t_{\rm visc} \simlt$ a few million years ($\approx 10^6$ years
at 0.2''), and we expect that gas would have mainly accreted
onto \sgra\ by now. This accretion rate would be a significant fraction of the
Eddington accretion rate. Standard disc accretion would then generate as much
as $2\times 10^{56}$ erg $M_{\rm acc, 3}$ of radiative energy, where $M_{\rm
acc, 3}$ is the gas mass accreted by \sgra\ in units of $10^3 \msun$. A
similar amount of energy could have been released as energetic outflows. There
is currently no evidence for such a bright and relatively recent period of accretion
activity for \sgra.

In the opposite limit, i.e. if $\alpha \ll 0.1$, and $t_{\rm visc} \simgt$ a
few million years, a gaseous disc should still be present as self-consistent
modelling predicts that stars should not be forming inside $R \simlt
0.3''-1''$ \citep{Nayakshin06a,Levin06}. There are however very strong
observational constraints on the absense of such a massive gaseous disc in the
inner arcsecond of our Galaxy \citep[e.g.,][]{Falcke97,Narayan02,Cuadra04}.

Therefore, our simulations seem to over-predict the amount of gaseous material
deposited in the inner $\sim 0.3''$. We take this as an indication that a
better model would perhaps involve a gaseous cloud of a larger geometric size, thus
shifting all spatial scales outwards (see further discussion of this in \S
\ref{sec:obs}).

\section{Discussion}\label{sec:discussion}

\subsection{Summary of main results}\label{sec:summary}

In terms of the final distribution of gas and resulting stellar populations,
our results can be divided into two main categories.
\begin{enumerate}
\item For those runs with a relatively small impact parameter, namely S1, S3
and S5, the initially small-scale disc around the black hole grows by steady
accumulation of gas to extend out to $R \sim 15''$. As a result, more than one
stellar population is seen in the disc. In addition to the ubiquitous
stars in the inner few arcseconds, a ``mid-range'' population is seen,
at $R \sim 5-8''$. Populations in the filaments vary: in S1, only the primary
filament forms a stellar population, as the secondary has accreted onto the
disc by the time stars begin to form; in S3, although the gas dynamics are the
same, the secondary filament forms stars before it can accrete, resulting in
two filamentary populations; and in S5, due to the small impact parameter a
single large disc is created, resulting in three disc stellar populations all
at a similar orientation.

\item For the runs with a larger impact parameter, namely S2, S4 and S6, the
small-scale disc grows very little over the course of the simulation, staying
within a radius of a few arcseconds. The feeding of the disc is far
from steady, generating enhanced midplane rotation, which, coupled with star
formation (particularly in the case of faster cooling, e.g., S4) results in
geometrically thick stellar populations. Both the primary and secondary
filaments survive, and each form stellar populations inclined at a large angle
to each other.
\end{enumerate}

The other results of our simulations, presented roughly in the order of their
commonality to the 6 simulations, are the following. Firstly, the formation
of a gaseous nearly circular disc in the inner region of the computational domain
is common to all runs, as is the ensuing formation of stars on similar
near-circular orbits. This is natural as the dynamical time in the innermost disc
is only $\sim 60 r^{3/2}$ years. On the scale of the inner disc, $r \sim 1$,
the disc makes tens to hundreds of revolutions during the simulation, allowing
for a near circularisation of the gaseous orbits.  Conversely, the outer
gaseous stream becomes self-gravitating much faster than it could circularise
(or even become an eccentric disc), and hence orbits of stars in that region
are more eccentric. This division on inner near-circular and outer eccentric orbits
is in broad agreement with the properties of the observed orbits
\citep{PaumardEtal06}. One discrepancy however is the eccentricity of the
counter-clockwise feature of the observations compared to the outer stellar
populations as seen in our model; the former is believed to possess a value of
$e \sim 0.6-0.8$ whilst the latter we find is only $e \sim 0.2-0.4$. We
believe that simulations that use a strongly eccentric initial orbit for
one of the clouds should be able to match the observations more closely.

Another robust result of our simulations is that all of the gaseous and
ensuing stellar discs are significantly warped, by between 30$^\circ$ to
60$^\circ$ measured from the inner to the outer radii. Interestingly, most
recent analysis of the observations reveals strong warps in the clockwise disc system
(Bartko et al 08, private communication). 

We also found that the radial distribution of stellar mass closely follows the
observed $\Sigma_* \propto 1/R^2$ profile of the disc stellar populations
\citep{PaumardEtal06}. This was observed for all the simulations, although we expect
these results would change if the colliding clouds moved in similar directions,
significantly reducing the angular momentum cancellation in the shock and the
thermal ``kick'' velocity due to the shock. Of course, this latter setup would
prevent the formation of stellar populations at a large angle of inclination with
respect to each other (one of the most prominent features of the observational data),
so it is reasonable that we have not tested this possibility in our model.

For the same initial configurations, simulations with a comparatively long cooling time
parameter, $\beta=1$, lead to kinematically less dispersed stellar populations
than those with faster cooling (i.e., $\beta=0.3$). As a result, the
longer the cooling time, the more closely the resulting stellar system can be
fit by planar systems in velocity space. The innermost stellar disc is then
reminiscent of the observed clockwise {\em thin} stellar system
\citep{PaumardEtal06} in terms of stellar orbits.  Rapid cooling produces
clumpier gas flows that lead to significant gaseous disc orientation changes
during the simulations. Rapidly cooling gas flows naturally form less coherent
discs; such geometrically thick systems are incompatible with the observations.

Interestingly, with the chosen initial conditions, faster cooling promotes
survival of the gaseous streams corresponding to the orbits of the original
clouds. These streams fragment and form stars mainly in a clustered mode,
although this might be expected to depend on the details of the radiation
feedback from young stars, which is not modelled in this set of simulations.

Our slower cooling runs produce inner and outer stellar systems that are
inclined to each other by only about $40^\circ$, which is much smaller than
the observed angle of $\sim 110^\circ$ between the discs
\citep{PaumardEtal06}. This is a natural consequence of angular momentum
conservation in a ``viscous'', well-mixed gas flow -- the disc orientation in
our model will always lie in-between the positions of the initial clouds in
angular momentum space.

\subsection{Origin of the observed stars: one or two clouds?}\label{sec:obs}

None of our runs reproduce the observed properties of stars and their orbits
in detail. This is not surprising as the parameter space for initial
conditions is quite large, and we have explored only a tiny fraction of it
here. We feel, however, that the agreement with the observations is
encouraging and that the model is promising. To quantify this discussion
here, we must note that the observational picture itself is still not entirely
clear. Whereas \cite{GenzelEtal03} and \cite{PaumardEtal06} classify the orbital stellar
distribution as two discs, the statistical significance of the second disc is
disputed by \cite{LuEtal06,LuEtal08}. Although we tend to agree with the first
opinion, we shall consider both of these possibilities without bias here.

Let us first discuss the origin of the clockwise stellar disc in the inner
parsec of the GC. The existence of this disc is not in question since the work
of \cite{Levin03} and \cite{GenzelEtal03}. Concentrating on this feature {\em only}, a
viable model can be simplified to a self-collision of a large molecular cloud
rather than collision of two clouds. Recently, \cite{WardleYZ08}, argued
that a viable progenitor for the stellar discs is a large-scale molecular
cloud, such as the $50$ km s$^{-1}$ GMC actually observed near the GC. In this
model, the cloud infalls almost radially onto \sgra, temporarily engulfing the
SMBH. While most of the cloud nearly passes by, gas within one parsec or so
gets captured by the SMBH and settles into a bound disc of gas. This model may
be also useful in explaining not only the observed stellar disc within the
inner parsec but also the CND further out at $\sim 10$ pc. 

Even more recently, \cite{Bonnell08} have performed numerical simulations of a scenario similar to
that suggested by \cite{WardleYZ08}. In particular, an infalling turbulent
molecular cloud was set on a sub-parsec impact parameter with respect to the
SMBH.  While their numerical method did not include radiative transfer
directly (it is of course also not included here), they did use an effective
equation of state that to some degree mimics the radiative effects. It was
found that the cloud becomes tidally disrupted and forms an eccentric disc
around the SMBH. This disc later fragmented into stars with a top-heavy IMF
resulting from inefficient cooling. Therefore, the one cloud scenario
appears to be attractive both qualitatively and quantitatively as an explanation for the
origin of the clockwise stellar disc.

Now, considering the rest of the young massive stars that do {\em not} belong
to the clockwise disc kinematically, we note that these account for no less
than 50\% of the overall stellar population. Whether the majority of these
stars form the second disc or not, their velocity vectors differ from the
clockwise disc by up to a few hundred km/sec (which is the order of circular
velocity at these distances). The velocity dispersion even in a very compact
Giant Molecular Cloud, with mass of say $3\times 10^4 \msun$ and size of 1 parsec,
is only $\sim 10$ km/sec. Thus the only way to create the observed
kinematically distinct population of stars would be to postulate the existence of
two or more streams (filaments) inside the cloud that pass on opposite sides
of \sgra\ and do {\em not} get completely mixed before forming stars. Given our
numerical experiments in this paper, this does not seem implausible if the cooling
time is short $\beta \simlt 1$. What is interesting in this scenario
is that the massive stars of the counter-clockwise population would then have to
form very quickly, i.e., on a dynamical timescale, or else gaseous orbits
would be mixed. The rotation period scales approximately as $T_{\rm rot} =
3000 (R/5'')^{3/2}$ years, so this is quite fast indeed. 

It should also be noted that recent N-body simulations \citep{Cuadra08, Alexander07} imply that it would have been very difficult for the high
eccentricities and inclinations of the dynamically hotter counter-clockwise feature to have been formed from a flat, cold disc
via scattering processes. A single disc progenitor for both GC stellar features is therefore largely ruled out. In contrast, in the case of the collision of two clouds as considered here, it is
almost too easy to obtain an inner near-circular disc and a kinematically diverse stellar
population farther out. We therefore favor a model where a GMC collided
with a pre-existing cloud or structure, such as a massive larger-scale disc,
e.g., similar to the observed CND.

\subsection{Size of the cloud(s)}\label{sec:cloud_size}

The observed well defined, flat, geometrically thin and near-circular/mildly eccentric
clockwise stellar system \citep{PaumardEtal06} is best created via a gentle
accumulation of gas. Several independent major gas deposition events lead to a
warped disc, and/or mixed systems consisting of several stellar rings
or discs co-existing at the same radius. To avoid this happening, the inner
disc must be created on time scale longer than the critical rotation time,
which is estimated at $t_{\rm cr} \sim$~few~$\times 10^4$~years.

While these results are based on analytical arguments \citep{NC05} and our simplified
one-parameter $\beta$-cooling model, we note that the results of \cite{Bonnell08}
corroborate this as their inner stellar disc appears to be too eccentric to
match the data of \cite{PaumardEtal06}.

Deposition of gas in the inner disc takes place on the longest of two
timescales: the cooling time $t_{\rm cool}$ and the collision time, $t_{\rm
coll} \sim R_{\rm cl}/v_{\rm cl}$, where $R_{\rm cl}$ and $v_{\rm cl}$ are the
cloud's size and velocity magnitude. In the appendix we estimate the
realistic cloud cooling time {\em during the collision}, and show that it is
always much shorter than the dynamical time unless magnetic fields are very
important. 

We are thus left with the only option to require the collision itself be more
prolonged than $t_{\rm cr}$. Estimating the velocity of the cloud at $v_{\rm cl}
\sim 150$ km/sec, which is of order of circular velocities in the inner Galaxy
outside the inner parsec, we find $t_{\rm coll} = R_{\rm cl}/v_{\rm cl} \sim
10^4 \hbox{years}\; R_{\rm cl, pc}$, where $R_{\rm cl, pc}$ is the size of the
cloud in parsecs. We hence require the cloud to be larger than a few parsecs to
satisfy $t_{\rm coll}\simgt t_{\rm cr}$. Note that this size is not
necessarily the original size of the cloud if the cloud gets tidally disrupted
before it makes the impact. In the latter case we can take $R_{\rm cl}$ to be
the radial distance to the centre of the Galaxy at which the tidal disruption
took place. Finally, the location of the collision should not be too far from
the central parsec, or else too much angular momentum would have to be lost to
deposit a significant amount of gas at $\sim 0.1$ pc. 

Another argument going in the same direction comes from a comparison of the
radial distribution of gas and stars in our simulations with the observed
stellar distribution \citep{PaumardEtal06}. The former is too compact, i.e.,
all of our simulations deposited {\em too much} mass within the inner arcsecond. In
addition, if that was indeed the case 6 million years ago, then \sgra\ would have
received a significant amount of fuel, enough to become at least a bright AGN. Given
the long viscous times in the inner arcsecond, \sgra\ could actually continue
to accrete this fuel now. However, it is well known that there is no
geometrically thin and optically thick disc inside the inner arcsecond of
\sgra\ \citep{Falcke97,Narayan02,Cuadra04}. Eliminating the gaseous disc by
star formation is not an option as there are not enough massive young stars
observed there.

Taking all these constraints together, we believe that the most realistic
scenario would be a GMC of the order of a few parsecs in size
striking the CND at the distance of a few parsecs from
\sgra. This scenario could perhaps explain the origin of the inner edge of the
CND at $R\approx2$~pc if the refilling time scale is longer than the age of
the young stars. Alternatively such a cloud could self-collide if the impact
parameter with respect to \sgra\ is small enough, but the cloud needs to be
very structured, e.g., essentially consist of several smaller clouds or
filaments.

\section{Conclusions}

In this paper we presented several simulations of cloud-cloud collisions aimed
at reproducing gas flows that could have formed (a) gaseous disc(s) in the central
parsec of our Galaxy, as well as the resulting star formation. We found the
gas cooling time and the impact parameter of the collision to influence the
outcome significantly. Nevertheless, there are several robust results: (a)
the inner near-circular and outer eccentric orbital structure of the stars formed
there; (b) a sharply peaked mass distribution of stars $\Sigma_*(R) \sim
1/R^2$; (c) that the gaseous and stellar discs are warped. These results are in
good accordance with the observations. The breakdown of the stellar system into one
or more components is sensitive to the initial conditions and also the
cooling parameter.

It appears that a GMC with a size of one to a few parsecs, self-colliding on a
nearly radial orbit, or striking the CND at the distance of a few parsec from
\sgra\ could explain the known observational data satisfactorily. Future
observations and modelling might put interesting upper limits on
the timescale over which the massive stars formed in the GC, and detail the
structure and orbit of such a GMC.

\section{Acknowledgments}

We acknowledge useful discussions by Peter Cossins, Giuseppe Lodato and
Yuri Levin, and we thank the referee for their comments on the draft.

\bibliographystyle{mnras} 

\bibliography{nayakshinalex}

\appendix

\section{On realistic estimates of the cooling
  time during the cloud collision}\label{sec:appendix}

Here we try to estimate the realistic cooling time for the gas clouds in the
initial collision event. Clouds moving at a few
hundred km/sec would yield a maximum shock temperature of a few million
K. Using the optically thin cooling function at these temperatures (dominated
by Bremsstrahlung and metal line cooling) and initial cloud densities,
one finds that the optically thin cooling time is $3-5$ orders of magnitude
shorter than dynamical time at $R \sim 25''$, the latter being $t_{\rm dyn} = 8 \times 10^3$ yrs.
With our chosen parameters for the initial conditions, the clouds are Compton-thick, 
i.e. $\Sigma \sim $ a few tens of g cm$^{-2}$. The shock would emit soft X-rays through 
Bremsstrahlung, which would have an absorption opacity much higher than that of 
the Thompson opacity i.e. $\kappa_{\rm T} = 0.4$ cm$^2$ g$^{-1}$. These X-rays would
therefore be completely absorbed within the colliding clouds. This would not
lead to a significant increase in the cooling time, however, as the absorbed
X-rays would be re-emitted as thermalised blackbody emission at the effective
temperature of the radiation. The latter is estimated to be around $10^3$
K. The clouds' optical depth to such radiation is of the order of a few tens, meaning that
the clouds would still be able to cool effectively via blackbody emission.
The cooling time for the cloud in its entirety would therefore
remain short (using the Stefan-Boltzmann law for radiative emission,
$t_{\rm cool}$ would again be $\sim 3$ orders of magnitude shorter than
$t_{\rm dyn}$).

However, as was pointed out to us by Yuri Levin, these considerations
completely neglect magnetic fields that are most likely present in the
pre-collision clouds. If the radiative cooling time is short, then the shock
is essentially isothermal, and we can expect the gas to be compressed in the
shock by a few orders of magnitude for our typical conditions. A frozen-in
magnetic field would then be amplified by similar factors, and magnetic
pressure by four to six orders of magnitude. The strongly increased magnetic
pressure might present another form of energy and pressure support against
gravitational collapse. If magnetic flux is then removed from the cloud by
magnetic buoyancy instability, this happens on time scales longer than
dynamical, effectively implying large values of $\beta$. 

Summarising, a realistic value of $\beta$ during the initial cloud-cloud collision would be
very small if magnetic fields are not important, or large if strong field
amplification occurs due to the shock.

\end{document}